\documentclass[11pt,a4paper]{article}
\pdfoutput=1   % required for arXiv submission if pdflatex is used
\usepackage[utf8]{inputenc}
\usepackage{tabularx}
\usepackage{lmodern}
\usepackage[outercaption]{sidecap}
\usepackage{macros}
\usepackage{xspace}
\usepackage{cancel}
\usepackage{enumerate}
\usepackage{booktabs}
\usepackage{mathtools}
\usepackage{alpha}
\hypersetup{%
   pdftitle    = {},
   pdfauthor   = {G.M.~de\,Divitiis, P.~Fritzsch, J.~Heitger, C.C.~K\"oster, S.~Kuberski, A.~Vladikas},
   pdfkeywords = {Lattice QCD, Non-perturbative Effects, Symanzik Improvement}
}
\usepackage[colorinlistoftodos,prependcaption,textsize=tiny]{todonotes}
\newcommand\setItemnumber[1]{\setcounter{enumi}{\numexpr#1-1\relax}}
\newcolumntype{C}{>{$}c<{$}}% whole column in mathmode => no $ expected !!!
\newcolumntype{R}{>{$}r<{$}}% whole column in mathmode => no $ expected !!!
\newcolumntype{L}{>{$}l<{$}}% whole column in mathmode => no $ expected !!!

\begin{document}

\preprintno{%
CERN-TH-2019-085\\
MS-TP-19-13
\vfill
}

\title{%
Non-perturbative determination of improvement coefficients $\bm$ and $\ba-\bp$
and normalisation factor $\zm \zp/\za$ with $\NF = 3$ Wilson fermions
}

\collaboration{\includegraphics[width=2.8cm]{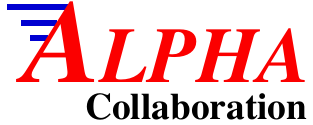}}

\author[utv,infn]{Giulia~Maria de~Divitiis}
\author[cern]{Patrick Fritzsch}
\author[ms]{Jochen Heitger}
\author[ms]{Carl~Christian K\"oster}
\author[ms]{Simon Kuberski}
\author[infn]{Anastassios Vladikas}

\address[utv]{Universit\`{a} di Roma ``Tor Vergata'', Dipartimento di Fisica, \\ Via della Ricerca Scientifica~1, 00133 Rome, Italy}
\address[infn]{INFN, ``Rome Tor Vergata'' Division, c/o Dipartimento di Fisica, \\ Via della Ricerca Scientifica~1, 00133 Rome, Italy}
\address[cern]{Theoretical Physics Department, CERN, 1211 Geneva 23, Switzerland}
\address[ms]{Westf\"alische Wilhelms-Universit\"at M\"unster, Institut f\"ur Theoretische Physik,\\ Wilhelm-Klemm-Stra{\ss}e 9, 48149 M\"unster, Germany}

\begin{abstract}
We determine non-perturbatively the normalisation parameter $\zm\zp/\za$ as
well as the Symanzik coefficients $\bm$ and $\ba-\bp$, required in
${\rm O}(a)$ improved quark mass renormalisation with Wilson fermions. 
The strategy underlying their computation involves simulations in $\NF=3$ 
QCD with ${\rmO}(a)$ improved massless sea and non-degenerate valence 
quarks in the finite-volume Schr\"odinger functional scheme.
Our results, which cover the typical gauge coupling range of large-volume
$\NF=2+1$ QCD simulations with Wilson fermions at lattice spacings below
$0.1\,\Fm$, are of particular use for the non-perturbative calculation of
${\rm O}(a)$ improved renormalised quark masses.\\
\end{abstract}


\maketitle

\tableofcontents

% place here, otherwise minor issues with title page appear until fixed
\makeatletter
\g@addto@macro\bfseries{\boldmath}
\makeatother

\section{Introduction}
\label{sec:intro}

Quark masses are amongst the fundamental parameters of the theory of strong
interactions. Their high-precision determination is one of the main goals of
lattice QCD (see ref.~\cite{Aoki:2019cca} and references therein). These
computations suffer from statistical and systematic uncertainties, which can be
reduced in a controlled way. An important source of uncertainty are cutoff
effects, which are removed by computing a given quantity at several lattice
spacings, followed by continuum extrapolation.
For several variants of lattice fermions
(staggered, domain wall, overlap, twisted-mass) these uncertainties are
$\rmO(a^2)$, while for Wilson fermions they are $\rmO(a)$. A related problem in
the latter formulation is the loss of chiral symmetry, because it complicates
the renormalisation properties of most quantities. A frequently cited example
of these complications is the power divergence $\m_{\rm crit} \sim 1/a$ that
must be subtracted from bare quark masses before they are renormalised
multiplicatively. Another example is the fact that the normalisation factor
$Z_{\rm A}$ of the axial current and the ratio $Z_{\rm S}/Z_{\rm P}$ of the
scalar and pseudoscalar density renormalisation parameters are finite functions
of the gauge coupling, which are equal to unity only in the continuum limit
where chiral symmetry is fully recovered.

In spite of these shortcomings, Wilson fermions have advantages compared to
other popular regularisations, namely strict locality (leading to relatively
reduced computational costs) and preservation of flavour symmetry. 
It is the regularisation of choice of our collaboration, which is part of the
effort by the CLS (Coordinated Lattice Simulations) cooperation to simulate
QCD with $\nf=2+1$ flavours of non-perturbatively improved Wilson
fermions~\cite{Bruno:2014jqa,Bruno:2016plf,Bali:2016umi,Mohler:2017wnb}.

Wilson fermion $\rmO(a)$ discretisation effects are systematically removed by 
introducing so-called Symanzik counter-terms in the lattice action and
composite operators. These counter-terms are higher dimensional operators with
coefficients which are functions of the gauge coupling. The coefficients must
be appropriately tuned so that $\rmO(a)$ improvement is achieved. Some of them
($\csw, \ca$, etc.) remove discretisation effects which are present
also in the chiral limit, whereas others ($\bm, \ba, \bp$, etc.) are
proportional to the quark masses and improve quantities off the chiral limit.
The requirement for improvement in the fermionic sector has been noted early 
on~\cite{Heatlie:1990kg}, and only a few strategies to determine them 
non-perturbatively have been developed so far.

In the present work we compute non-perturbatively the coefficients
$\bm$, $\ba - \bp$ and the renormalisation parameter
$Z\equiv Z_{\rm m} Z_{\rm P}/Z_{\rm A}$ in a theory of three sea quark flavours.
The methods we use can be traced back to
ref.~\cite{deDivitiis:1997ka}. In that work, renormalised quark masses were
defined both through the PCAC bare quark masses and the subtracted bare Wilson
masses. In both definitions $\rmO(a)$ improvement is introduced through the
inclusion of all necessary $c$- and $b$-type counter-terms. Combining these
results at constant bare gauge coupling provides estimates of $\bm$, $\ba -
\bp$ and $Z$. This work has been extended in
refs.~\cite{Bhattacharya:1999uq,Bhattacharya:2000pn,Bhattacharya:2005ss}, where
results for other improvement coefficients were also reported. These
computations were performed in large volumes with (anti)periodic boundary
conditions. In parallel, in ref.~\cite{Guagnelli:2000jw} (and subsequently
in~\cite{HQET:pap2}) the method was extended and applied to small physical
volumes with Schr\"odinger functional
boundary conditions. These early analyses were carried out in the quenched
approximation. More recently, in ref.~\cite{Fritzsch:2010aw}, $\bm$,
${\ba-\bp}$ and $Z$ were measured in a theory with $\NF=2$ sea quarks,
employing the
Schr\"odinger functional scheme and working at a constant value of the
renormalised coup\-ling, so as to keep the physical extent of the lattice fixed.
By thus imposing improvement and renormalisation conditions along a line in
lattice parameter space, where all physical scales stay constant, it is
ensured that any intrinsic higher-order lattice spacing ambiguities of
$b$-coefficients and $Z$-factors vanish uniformly as the continuum limit is
approached.

Our strategy follows closely that of ref.~\cite{Fritzsch:2010aw}. However, the
extraction of the final estimates from our data has been improved by the
introduction of several novelties in the data analysis, which enable us to
obtain very reliable estimates in the chiral limit. The lattice action we
employ consists of the tree-level Symanzik-improved gauge
action~\cite{Luscher:1984xn} and the non-perturbatively improved Wilson-clover
fermion action~\cite{Sheikholeslami:1985ij}. Our simulations are performed in
the range of bare couplings, where gauge configurations on lattices with large
physical volumes with $\NF=2+1$ sea quarks have been generated by
CLS~\cite{Bruno:2014jqa,Bali:2016umi}.
These configurations are suitable for the computation
of bare correlation functions, on the basis of which low-energy hadronic
quantities can be evaluated. In ref.~\cite{Bruno:2016plf} bare PCAC quark
masses have been computed from these ensembles. To obtain renormalised up,
down, and strange quark masses from these bare masses, one also needs the 
following: 
(i) The multiplicative mass renormalisation factor $1/Z_{\rm P}$ at low energies
and its non-perturbative running up to high energy scales; these are known 
in the Schr\"odinger functional scheme from ref.~\cite{Campos:2018ahf}. 
(ii)~The axial current improvement coefficient $c_{\rm A}$ and its 
normalisation constant $Z_{\rm A}$, which are known from 
refs.~\cite{Bulava:2015bxa} and 
\cite{Bulava:2016ktf,DallaBrida:2018tpn}, respectively.
(iii) The improvement coefficient $\ba-\bp$, which is one of the
results of this work. (iv) The improvement coefficient
$\bar b_{\rm A}-\bar b_{\rm P}$, which is particularly difficult to estimate
but may be ignored, as it is sub-leading in perturbation theory.

Independent estimates of Symanzik $b$-coefficients, directly computed on CLS
ensembles and obtained with a variant of the coordinate space method of
ref.~\cite{Martinelli:1997zc}, have been reported in~\cite{Korcyl:2016ugy}.
A comparison of these results to ours may be found in
sect.~\ref{sec:results}.
Preliminary results of the present work have been reported in
ref.~\cite{deDivitiis:2017vvw}.
For recent determinations of $b$-coefficients in the vector channel of
three-flavour QCD with the same lattice action,
see also refs.~\cite{Fritzsch:2018zym,Gerardin:2018kpy}.
\section{Quark mass renormalisation and improvement with Wilson fermions}
\label{sec:renormimpr}

In this section we review the renormalisation and $\rmO (a)$ improvement of
quark masses in the framework of lattice regularisation with Wilson quarks.
These results were first derived in ref.~\cite{Luscher:1996sc} 
for QCD with degenerate masses
and generalised in ref.~\cite{Bhattacharya:2005rb}, which is
the basis of our r\'esum\'e. The starting point is the subtracted bare quark
mass of flavour $i$ ($i=1, \ldots , \Nf$),
\begin{align}\label{eq:bare-mass} 
\mqi \equiv m_{0,i} - m_{\rm crit} &= \dfrac{1}{2a} \left (\dfrac{1}{\kappa_i}-\dfrac{1}{\kappa_{\rm crit}} \right ) \;,
\end{align}
where $\kappa_i$ is the hopping parameter, $\kappa_{\rm crit}$ its critical
value corresponding to the chiral limit with $\Nf$ degenerate flavours, 
and $a$ is the lattice spacing.
In terms of the subtracted masses $\mqi$, the $\rmO(a)$ improved,
renormalised quark mass is given by
\begin{align}  \label{eq:mren-mq} 
   m_{i,\rm R} &= \zm \left\{ \left [\, \mqi  + (r_{\rm m} - 1)\frac{\Tr{\Mq}}{\Nf} \,\right ] + a B_{i} \right\}  +  \rmO(a^2) \;,  \\\notag
        B_{i} &= \bm \mqi^2 + \bar b_{\rm m} \mqi \Tr{\Mq} + (r_{\rm m} d_{\rm m} - \bm) \dfrac{\Tr{\Mq^2}}{\Nf}  +  (r_{\rm m} \bar d_{\rm m} - \bar b_{\rm m}) \dfrac{\Tr{\Mq}^2}{\Nf}  \;,
\end{align}   
where $\Mq = {\rm diag}(m_{{\rm q},1}, \ldots , m_{{\rm q},\Nf})$ is the $\Nf
\times \Nf$  bare mass matrix (of subtracted quark masses), and $B_{i}$ a
combination of Symanzik counter-terms cancelling $\rmO(a)$ mass-dependent
cutoff effects.
The $\Tr{\Mq}$-term in the square brackets appears at leading order and 
re\-presents a redefinition of the chiral point in the presence of massive 
quarks.

We recall in passing that the renormalisation parameter
$Z_\mathrm{m}(g_0^2,a\mu)$ depends on the renormalisation scale $\mu$ and
diverges logarithmically in the ultraviolet. A mass-independent renormalisation
scheme is implied throughout this work.  In such a scheme, the Symanzik
coefficients $\bm$, $\bar b_{\rm m}$, $d_{\rm m}$, $\bar d_{\rm m}$ as well as
$r_{\rm m}$ are functions of the squared bare coupling $g_0^2$.%
\footnote{%
In an improved mass-independent scheme, the coupling is to be defined as 
$\tilde g_0^2 \equiv g_0^2 (1+b_\mathrm{g}a\Tr{\Mq})$, 
see ref.~\cite{Luscher:1996sc}. As this coupling redefinition affects 
$b$-counter-terms by $\rmO(a^2)$ terms, we need not take it into 
consideration  in the present work.
}%
In a non-perturbative determination at non-zero quark mass, they are affected by
$\rmO(a\mqi)$ and $\rmO(a\Tr{\Mq})$ systematic effects, which are part of
their operational definition. They have the following properties:%
\footnote{%
The first three properties enumerated below are discussed in
ref.~\cite{Bhattacharya:2005rb}, while those of the last two are due
to S.R.~Sharpe (private communication).
}%
%
%% Figures     
%
\usetikzlibrary{decorations.markings}
\usetikzlibrary{decorations.pathmorphing}
%
%% Define Vertices    
%
\tikzset{mass vertex/.style={rectangle,draw=black,fill=black,inner sep=0pt, minimum size=0.14cm}}
\tikzset{vertex/.style={circle,draw=black,fill=black,inner sep=0pt, minimum size=0.04cm}}
%
%% Define Propagators   
%
\tikzset{->-/.style={decoration={
  markings,
  mark=at position #1 with {\arrow{>}}},>=stealth,postaction={decorate}}}
\tikzset{->>-/.style={decoration={
  markings,
  mark=at position #1 with {\arrow[black,line width=0.2mm]{>}}},postaction={decorate}}}
\tikzset{->m>-/.style={decoration={
  markings,
  mark=at position 0.3 with {\arrow[black,line width=0.2mm]{>}},
  mark=at position 0.5 with {\node[mass vertex] {};},
  mark=at position 0.8 with {\arrow[black,line width=0.2mm]{>}}},
  postaction={decorate}}}
\tikzset{->mm>-/.style={decoration={
  markings,
  mark=at position 0.3 with {\arrow[black,line width=0.2mm]{>}},
  mark=at position 0.4 with {\node[mass vertex] {};},
  mark=at position 0.525 with {\arrow[black,line width=0.2mm]{>}},
  mark=at position 0.6 with {\node[mass vertex] {};},
  mark=at position 0.8 with {\arrow[black,line width=0.2mm]{>}}},
  postaction={decorate}}}
\tikzset{quark/.style={->>-=0.5}}
\tikzset{quarkm/.style={->m>-}}
\tikzset{quarkmm/.style={->mm>-}}
\tikzset{gluon/.style={decorate,decoration={coil,aspect=0.8,amplitude=2pt, segment length=5pt}}}%
%
%% Draw Diagrams    
%
\newcommand{\diagi}{\vcenter{\hbox{
\begin{tikzpicture}[scale=1.0, transform shape]
\coordinate (e1) at (0.0,0.0);
\coordinate (e2) at (4.0,0.0);
\node[vertex] (v1) at (1.0,0.0) {};
\node[vertex] (v2) at (3.0,0.0) {};
\node[vertex] (v3) at (1.5,0.8) {};
\node[vertex] (v4) at (2.5,0.8) {};
\foreach \from/\to in {e1/v1, v1/v2, v2/e2}
              \draw[quark] (\from) -- (\to);
\draw [gluon,bend left=40] (v1) to (v3);
\draw [gluon,bend right=40] (v2) to (v4);
\draw[bend left=80,quarkm] (v3) to (v4);
\draw[bend left=80,quark] (v4) to (v3);
\end{tikzpicture}
}}}
\newcommand{\diagiia}{\vcenter{\hbox{
\begin{tikzpicture}[scale=1.0, transform shape]
\coordinate (e1) at (0.0,0.0);
\coordinate (e2) at (3.0,0.0);
\draw [quarkmm] (e1) to (e2);
\end{tikzpicture}
}}}
\newcommand{\diagiib}{\vcenter{\hbox{
\begin{tikzpicture}[scale=1.0, transform shape]
\coordinate (e1) at (0.0,0.0);
\coordinate (e2) at (4.0,0.0);
\node[vertex] (v1) at (1.0,0.0) {};
\node[vertex] (v2) at (3.0,0.0) {};
\node[vertex] (v3) at (1.5,0.8) {};
\node[vertex] (v4) at (2.5,0.8) {};
\foreach \from/\to in {e1/v1, v2/e2}
              \draw[quark] (\from) -- (\to);
\draw [quarkmm] (v1) to (v2);
\draw [gluon,bend left=40] (v1) to (v3);
\draw [gluon,bend right=40] (v2) to (v4);
\draw[bend right=80,quark] (v3) to (v4);
\draw[bend right=80,quark] (v4) to (v3);
\end{tikzpicture}
}}}
\newcommand{\diagiii}{\vcenter{\hbox{
\begin{tikzpicture}[scale=1.0, transform shape]
\coordinate (e1) at (0.0,0.0);
\coordinate (e2) at (4.0,0.0);
\node[vertex] (v1) at (1.0,0.0) {};
\node[vertex] (v2) at (3.0,0.0) {};
\node[vertex] (v3) at (1.5,0.8) {};
\node[vertex] (v4) at (2.5,0.8) {};
\foreach \from/\to in {e1/v1, v2/e2}
              \draw[quark] (\from) -- (\to);
\draw [quarkm] (v1) to (v2);
\draw [gluon,bend left=40] (v1) to (v3);
\draw [gluon,bend right=40] (v2) to (v4);
\draw[bend right=80,quarkm] (v3) to (v4);
\draw[bend right=80,quark] (v4) to (v3);
\end{tikzpicture}
}}}
\newcommand{\diagiv}{\vcenter{\hbox{
\begin{tikzpicture}[scale=1.0, transform shape]
\coordinate (e1) at (0.0,0.0);
\coordinate (e2) at (4.0,0.0);
\node[vertex] (v1) at (1.0,0.0) {};
\node[vertex] (v2) at (3.0,0.0) {};
\node[vertex] (v3) at (1.5,0.8) {};
\node[vertex] (v4) at (2.5,0.8) {};
\foreach \from/\to in {e1/v1, v1/v2, v2/e2}
              \draw[quark] (\from) -- (\to);
\draw [gluon,bend left=40] (v1) to (v3);
\draw [gluon,bend right=40] (v2) to (v4);
\draw[bend left=80,quarkm] (v3) to (v4);
\draw[bend left=80,quarkm] (v4) to (v3);
\end{tikzpicture}
}}}
\newcommand{\diagv}{\vcenter{\hbox{
\begin{tikzpicture}[scale=1.0, transform shape]
\coordinate (e1) at (0.0,0.0);
\coordinate (e2) at (4.0,0.0);
\node[vertex] (v1) at (1.0,0.0) {};
\node[vertex] (v2) at (3.0,0.0) {};
\node[vertex] (v3) at (1.2,0.7) {};
\node[vertex] (v4) at (1.7,0.9) {};
\node[vertex] (v5) at (2.3,0.9) {};
\node[vertex] (v6) at (2.8,0.7) {};
\foreach \from/\to in {e1/v1, v1/v2, v2/e2}
              \draw[quark] (\from) -- (\to);
\draw [gluon,bend left=40] (v1) to (v3);
\draw [gluon,bend right=40] (v2) to (v6);
\draw [gluon,bend left=20] (v4) to (v5);
\draw[bend right=80,quark] (v3) to (v4);
\draw[bend right=80,quarkm] (v4) to (v3);
\draw[bend right=80,quark] (v5) to (v6);
\draw[bend right=80,quarkm] (v6) to (v5);
\end{tikzpicture}
}}}
\newcommand{\diagvia}{\vcenter{\hbox{
\begin{tikzpicture}[scale=1.0, transform shape]
\coordinate (e1) at (0.0,0.0);
\coordinate (e2) at (3.0,0.0);
\draw [quarkm] (e1) to (e2);
\end{tikzpicture}
}}}
\newcommand{\diagvib}{\vcenter{\hbox{
\begin{tikzpicture}[scale=1.0, transform shape]
\coordinate (e1) at (0.0,0.0);
\coordinate (e2) at (4.0,0.0);
\node[vertex] (v1) at (1.0,0.0) {};
\node[vertex] (v2) at (3.0,0.0) {};
\node[vertex] (v3) at (1.5,0.8) {};
\node[vertex] (v4) at (2.5,0.8) {};
\foreach \from/\to in {e1/v1, v2/e2}
              \draw[quark] (\from) -- (\to);
\draw [quarkm] (v1) to (v2);
\draw [gluon,bend left=40] (v1) to (v3);
\draw [gluon,bend right=40] (v2) to (v4);
\draw[bend right=80,quark] (v3) to (v4);
\draw[bend right=80,quark] (v4) to (v3);
\end{tikzpicture}
}}}
\newcommand{\diagiic}{\vcenter{\hbox{
\begin{tikzpicture}[scale=1.0, transform shape]
\coordinate (e1) at (0.0,0.0);
\coordinate (e2) at (4.0,0.0);
\node[vertex] (v1) at (1.0,0.0) {};
\node[vertex] (v2) at (3.0,0.0) {};
\foreach \from/\to in {e1/v1, v2/e2}
              \draw[quark] (\from) -- (\to);
\draw [quarkmm] (v1) to (v2);
\draw [gluon,bend left=80] (v1) to (v2);
\end{tikzpicture}
}}}
\newcommand{\diagvic}{\vcenter{\hbox{
\begin{tikzpicture}[scale=1.0, transform shape]
\coordinate (e1) at (0.0,0.0);
\coordinate (e2) at (4.0,0.0);
\node[vertex] (v1) at (1.0,0.0) {};
\node[vertex] (v2) at (3.0,0.0) {};
\foreach \from/\to in {e1/v1, v2/e2}
              \draw[quark] (\from) -- (\to);
\draw [quarkm] (v1) to (v2);
\draw [gluon,bend left=80] (v1) to (v2);
\end{tikzpicture}
}}}
\newcommand{\diagvii}{\vcenter{\hbox{
\begin{tikzpicture}[scale=1.0, transform shape]
\coordinate (e1) at (0.0,0.0);
\coordinate (e2) at (4.0,0.0);
\node[vertex] (v1) at (1.0,0.0) {};
\node[vertex] (v2) at (3.0,0.0) {};
\node[vertex] (v3) at (1.5,0.8) {};
\node[vertex] (v4) at (2.5,0.8) {};
\foreach \from/\to in {e1/v1, v1/v2, v2/e2}
              \draw[quark] (\from) -- (\to);
\draw [gluon,bend left=40] (v1) to (v3);
\draw [gluon,bend right=40] (v2) to (v4);
\draw[bend left=80,quarkm] (v3) to (v4);
\draw[bend left=80,quark] (v4) to (v3);
\end{tikzpicture}
}}}

\begin{enumerate}[i]
\item [i)] The $(r_{\rm m}-1)$-term multiplies $\Tr{\Mq}$, so it arises from a
        mass insertion on a quark loop; i.e., from diagrams like
        $$\resizebox{3cm}{!}{$\diagi$},$$ where the filled square indicates a
        mass insertion. It is a two-loop effect, contributing at $\rmO(g_0^4)$.
        Its determination is beyond the scope of this paper.
\item [ii)] The $\bm$-term multiplies $\mqi^2$. Thus it arises from: (a) the
        mass dependence of valence quark propagators and (b) the
        mass-independent contributions of the fermion loops. The former
        dependence begins at tree-level, so that 
        $b_\mathrm{m}=-1/2+\rmO(g_0^2)$, corresponding to Feynman diagrams like
        $$\resizebox{3cm}{!}{$\diagiia$}\quad \mathrm{and} \quad
        \resizebox{3cm}{!}{$\diagiic$}.$$ The latter dependence begins at
        two loops, contributing at $\rmO( g_0^4)$; cf. for example diagram
        $$\resizebox{3cm}{!}{$\diagiib$}.$$ The quenched $\bm$-value differs
        from the one of the full $\Nf$ theory by the mass-indepen\-dent
        contributions of the fermion loops; consequently the difference arises
        at two loops and is $\rmO( g_0^4)$.
\item [iii)] The $\bar b_{\rm m}$-term multiplies $\mqi\Tr{\Mq}$. The factor of
        $\mqi$ comes from the valence line, while the $\Tr{\Mq}$ from a quark
        loop. It thus begins at two loops in perturbation theory ($\bar b_{\rm m}
        \sim \rmO( g_0^4)$) and vanishes in the quenched approximation;
        e.g., diagram $$\resizebox{3cm}{!}{$\diagiii$}.$$ The determination of
        the $\bar b_{\rm m}$-term is beyond the scope of this paper.
\item [iv)] The $(r_{\rm m} d_{\rm m} - \bm)$-term multiplies $\Tr{\Mq^2}$; so
        it must arise from two insertions of $\mqi$ on a single sea-quark loop.
        This combination begins at two loops, so it is $\rmO( g_0^4)$; cf.
        diagram $$\resizebox{3cm}{!}{$\diagiv$}.$$ But since it arises from sea
        quark propagators, while $\bm$ has tree-level and $\rmO( g_0^2)$
        valence-quark contributions, it follows that also $d_{\rm m}$ must get
        tree-level contributions and $\rmO( g_0^2)$ corrections from
        valence lines. The determination of $(r_{\rm m} d_{\rm m} - \bm)$ is
        beyond the scope of this paper.
\item [v)] The $(r_{\rm m} \bar d_{\rm m} - \bar b_{\rm m})$-term multiplies
        $\Tr{\Mq}^2$, so it can only arise from mass insertions on two separate
        quark loops. Thus this term begins at three loops and is $\rmO(
        g_0^6)$; cf. diagram  $$\resizebox{3cm}{!}{$\diagv$}.$$ But since $\bar
        b_{\rm m}$ itself begins at two loops, so must $\bar d_{\rm m}$. The
        determination of $(r_{\rm m} \bar d_{\rm m} - \bar b_{\rm m})$ is
        beyond the scope of this paper.
\end{enumerate}

Next we recall that with Wilson fermions the renormalised quark mass can also
be related to the bare PCAC mass $m_{ij}$, {\it defined} through the following
relation:
\begin{equation}
(\partial_\mu)_x\evalbig{(\aimpr)^{ij}_\mu(x)\,\mathcal{O}^{ji}} \,=\, 2\,m_{ij} \, {\evalbig{{P}^{ij}(x)\,\mathcal{O}^{ji}}} \;,
\label{eqn:PCAC-mass}
\end{equation}
where $m_{ij} = \left(m_{i}+m_{j}\right)/2$.
Our notation is standard: 
the non-singlet bare axial current and the pseudoscalar density are given by
\begin{equation}
A_{\mu}^{ij}(x) \,\equiv\, \psibar_i(x)\,\dirac\mu\dirac5\,\psi_j(x) \,\, , \qquad 
P^{ij}(x) \,\equiv\, \psibar_i(x)\,\dirac5\,\psi_j(x) \,\, ,
\end{equation}
with indices $i,j$ denoting two distinct flavours. The pseudoscalar density
$P^{ij}$ and the current $(\aimpr)^{ij}_\mu \equiv A^{ij}_\mu+a\ca\partial_\mu
P^{ij}$ are Symanzik-improved in the chiral limit, with the improvement
coefficient $\ca(g_0^2)$ being in principle only a function of the gauge
coupling.%
\footnote{%
To be precise, for the divergence of the improved axial current we use  
${\partial_\mu (\aimpr)^{ij}_\mu \equiv \sdrv\mu A^{ij}_\mu
+a\ca \partial_\mu^\ast\partial_\mu P^{ij}}$,
where $\sdrv\mu$~denotes the average of the usual forward and backward 
derivatives defined as
${a\partial_\mu f(x) \equiv f(x+a\hat\mu) - f(x)}$ and 
${a\partial_\mu^\ast f(x) \equiv f(x) - f(x-a\hat\mu)}$.
}%

The operator
$\mathcal{O}^{ji}$ is defined in a region of space-time that does not include
the point $x$, thus avoiding contact terms. Our specific choice of correlation 
functions for eq.~(\ref{eqn:PCAC-mass}) is discussed in appendix~\ref{app:SF}.

Beyond the chiral limit, composite operators require improvement through the
introduction of $b$-type Symanzik counter-terms.  The renormalised and
$\rmO(a)$ improved axial current and pseudoscalar density are given
by~\cite{Luscher:1996sc,Bhattacharya:2005rb}
\begin{align}
        (\ar)_{\mu}^{ij}(x) &= \za(\gosq)\left[\,1+\ba(g_0^2)\,a m_{{\rm q},ij} + \bar b_{\rm A}(g_0^2)\,a \Tr{\Mq} \,\right](A_{\rm I})_{\mu}^{ij}(x) \;, 
\label{eqn:ren-off-diag-A} \\[0.5em]
(\pr)^{ij}(x) &= \zp(\gosq,a\mu)\left[\,1+\bp(g_0^2) \,a m_{{\rm q},ij} + \bar b_{\rm P}(g_0^2)\,a \Tr{\Mq} \, \,\right]P^{ij}(x) \,\, ,
\label{eqn:ren-off-diag-P}
\end{align}
with $m_{{\rm q},ij} \equiv \left(\mqi+\mqj\right)/2$.  The normalisation of
the axial current $Z_\mathrm{A}(g_0^2)$ is scale-independent, depending only on
the squared gauge coupling $g_0^2$.  The renormalisation parameter
$Z_\mathrm{P}(g_0^2,a\mu)$ also depends on the renormalisation scale $\mu$ and
diverges logarithmically in the ultraviolet. The Symanzik coefficients $\ba$,
$\bp$, $\bar b_{\rm A}$ and $\bar b_{\rm P}$ are in principle only functions of
the bare squared coupling. They have the following properties:
\begin{enumerate}[i]
\setItemnumber{6}
\item [vi)] The $\ba$- and $\bp$-terms multiply $m_{{\rm q},ij}$. Thus they
        arise from: (a) the mass dependence of valence quark propagators and
        (b) the mass-independent contributions of the fermion loops. The former
        dependence begins at tree-level, so that $\ba, \bp \, = \, 1/2 +
        \rmO( g_0^2)$; cf. diagrams $$\resizebox{3cm}{!}{$\diagvia$}\quad
        \mathrm{and} \quad\resizebox{3cm}{!}{$\diagvic$}.$$ The latter
        dependence begins at two loops, contributing at $\rmO( g_0^4)$; cf.
        diagram $$\resizebox{3cm}{!}{$\diagvib$}.$$ The difference $\ba - \bp$,
        appearing in eq.~(\ref{eq:renmassPCAC}) below, is therefore
        $\rmO( g_0^2)$.  The quenched values differ from those of the
        full $\Nf$ theory by the mass-independent contributions of the fermion
        loops; consequently the difference arises at two loops and is
        $\rmO( g_0^4)$.
\item [vii)] The $\bar b_{\rm A}$- and $\bar b_{\rm P}$-terms multiply
        $\Tr{\Mq}$. They arise from the mass dependence of quark fermion loops.
        They begin at two loops in perturbation theory ($\bar b_{\rm A}, \bar
        b_{\rm P} \sim \rmO( g_0^4)$) and vanish in the quenched
        approximation; cf. diagram $$\resizebox{3cm}{!}{$\diagi$}.$$ The
        determination of these coefficients is beyond the scope of this paper.
\end{enumerate}

The renormalised PCAC relation
\begin{align}
\evalbig{\, \partial_\mu (\ar)^{ij}_\mu(x)\ \mathcal{O}^{ji}\, } &=  
(\mri{i}+\mri{j})\,\evalbig{\, (P_\mathrm{R})^{ij}(x)\ \mathcal{O}^{ji}\,}+\Or(a^2) \,\, ,
\label{eqn:renorm-PCAC-relation-pm}
\end{align}
valid up to $\rmO(a^2)$ effects, combined with
eqs.~(\ref{eqn:PCAC-mass})--(\ref{eqn:ren-off-diag-P}), implies that
\begin{equation}
\label{eq:renmassPCAC}
\dfrac{m_{i,\rm R} + m_{j,\rm R}}{2} \,=\, \dfrac{\za}{\zp} \, m_{ij}
\Big [\, 1 \, + \, (\ba-\bp) a m_{{\rm q},ij} \, + \, (\bar b_{\rm A} - \bar b_{\rm P})a \Tr{\Mq} \,\Big ] \,+\, \rmO(a^2) \,\, .
\end{equation}

The properties of the various $b$-coefficients listed above 
eqs.~(\ref{eq:mren-mq}), (\ref{eqn:ren-off-diag-A}), (\ref{eqn:ren-off-diag-P})
and (\ref{eq:renmassPCAC}) also apply to the non-unitary theory, where
valence and sea quarks of the same flavour have different masses. We saw that
all terms containing traces of the fermion matrix refer to sea quarks, while
the others refer to valence quarks. This is shown somewhat more explicitly in
appendix~\ref{app:pQCD}. The present and previous works, such as
ref.~\cite{Fritzsch:2010aw}, rely on this property in order to obtain reliable
non-perturbative estimates of the Symanzik coefficients $\bm$ and $\ba-\bp$, 
as well as the combination
\begin{equation}
Z \,\equiv\, \zm\,\frac{ Z_{\rm P}}{Z_{\rm A}} \,\, . 
\end{equation}

\subsection{Non-perturbative determination of $\bm$, $\ba-\bp$ and $Z$}
\label{subsec:strategy}

If we calculate the average mass $(m_{i,\rm R} + m_{j,\rm R})/2$ from
eq.~(\ref{eq:mren-mq}) and equate the result to the r.h.s. of
eq.~(\ref{eq:renmassPCAC}), we obtain an expression, which relates subtracted
and PCAC bare masses:
\begin{align}\label{eq:massmatch}
    m_{ij}  &=  \dfrac{\zm \zp}{\za} \left\{ \left[\, \mqn{ij} + (r_{\rm m}-1) \dfrac{\Tr{\Mq}}{\NF} \,\right]  + aB_{ij} \,\right \}   +  \rmO(a^2) \;,  \\
    B_{ij  }&=  \bm \frac{\mqn{i}^2+\mqn{j}^2}{2}  -  (\ba-\bp)\mqn{ij}^2                                                                                       \nonumber \\
            &\quad + \Big ( \bar b_{\rm m}  - (\ba-\bp)  \dfrac{(r_{\rm m}-1)}{\NF}  -  (\bar b_{\rm A} - \bar b_{\rm P}) \Big ) \mqn{ij}  \Tr{\Mq}   \nonumber \\
            &\quad + (r_{\rm m} d_{\rm m}   - \bm) \dfrac{\Tr{\Mq^2}}{\NF} + \Big ( (r_{\rm m} \bar d_{\rm m} - \bar b_{\rm m}) - (r_{\rm m}-1) (\bar b_{\rm A} - \bar b_{\rm P}) \Big ) \dfrac{\Tr{\Mq}^2}{\NF}  \;. \notag
\end{align} 
Note that the product of the renormalisation parameters
$Z_\mathrm{P}(g_0^2,\mu) Z_\mathrm{m}(g_0^2,\mu)$ is scale independent.

As discussed previously and in appendix~\ref{app:pQCD},
eq.~(\ref{eq:massmatch}) remains valid in the non-unitary theory, with $\mqi$
denoting valence quark masses and $\Mq$ the mass matrix of sea quarks. Since
$\ba-\bp$, $\bm$ and the combination $Z \equiv \zm Z_{\rm P}/Z_{\rm A}$ are
short distance quantities, they can be determined in small physical volumes
with Schr\"odinger functional boundary conditions. This allows for simulations
with degenerate sea quarks lying very close to the chiral limit, so that terms
containing $\Tr{\Mq}$ can be dropped in eq.~(\ref{eq:massmatch}). Following the
strategy already proposed in refs.~\cite{Guagnelli:2000jw,Fritzsch:2010aw}, we
introduce two valence quark flavours with subtracted masses $m_{{\rm q},1} <
m_{{\rm q},2}$ and their average $m_{{\rm q},3} \equiv (m_{{\rm q},1} +m_{{\rm
q},2})/2$.  It is then straightforward to obtain estimators of the desired
improvement coefficients and normalisation factor $Z$ from the ratios
\begin{subequations}\label{eqn:estim-RX}
\begin{align}\label{eqn:estim-RAP}
\RAP &\equiv \dfrac{2\left(2\mij{12}-\mij{11}-\mij{22}\right)}{\left(\mij{11}-\mij{22}\right)\left(a\mqn{1}-a\mqn{2}\right)} = (\ba-\bP)\big\{ 1+\Or\big(a\mqn{12}; a \Tr{\Mq} \big)\big\} \;,
 \\[0.4em]\label{eqn:estim-Rm}
\Rm &\equiv \dfrac{4\left(\mij{12}-\mij{33}\right)} {\left(\mij{11}-\mij{22}\right)\left(a\mqn{1}-a\mqn{2}\right)} = \bm \big\{ 1+\Or\big(a\mqn{12} ; a \Tr{\Mq}  \big)\big\}  \;,
 \\[0.4em]\label{eqn:estim-RZ}
R_{Z} &\equiv \dfrac{\mij{11}-\mij{22}}{\mqn{1}-\mqn{2}} +\left(\RAP-\Rm\right)\left(a\mij{11}+a\mij{22}\right) = Z \big\{ 1 + \Or\big(a^2; a \Tr{\Mq}\big) \big\} \;,
\end{align}
\end{subequations}
where bare PCAC masses $\mij{ii}$ (with $i=1,2,3$) are defined through
eq.~(\ref{eqn:PCAC-mass}) for two degenerate but distinct flavours.

Note that the leading improvement coefficients obtained from $\RAP$ and $\Rm$
suffer from mass-dependent
$\Or(a)$ effects, which introduce only $\rmO(a^2)$ uncertainties in
the quark masses; cf. eqs.~(\ref{eq:mren-mq}) and (\ref{eq:renmassPCAC}). The
$\Or(a \Tr{\Mq})$ effects appearing on the r.h.s. of eqs.~(\ref{eqn:estim-RAP}),
(\ref{eqn:estim-Rm}) and (\ref{eqn:estim-RZ}) arise from the presence of a
residual non-zero sea quark mass in realistic simulations. This uncertainty is
removed once simulations are performed for several sea quark masses and the
chiral limit is reached by extrapolation. In our setup we are also able to 
simulate negative current quark masses, so in practice interpolation to the 
chiral limit is also possible.
In the chiral limit, the leading
normalisation factor $Z$ of the estimator $R_Z$ suffers from $\Or\big(a^2\big)$
effects; this is easily derived from eqs.~(\ref{eq:massmatch}) and
(\ref{eqn:estim-RZ}).

The above ratios  are not the only possible estimators of the quantities of
interest. For example, we can modify $\RAP$ and $\Rm$ by replacing the
denominator $\left(\mij{11}-\mij{22}\right)$ by any of the following PCAC mass
differences: $2(\mij{22}-\mij{33})$, $2(\mij{33}-\mij{11})$,
$2(\mij{22}-\mij{12})$, or $2 (\mij{12}-\mij{11})$. As shown in
ref.~\cite{deDivitiis:2017vvw}, these new ratios also provide estimates of
$(\ba - \bp)$, $\bm$ and $Z$, with different finite cutoff effects. In
practice these differences were found to be orders of magnitude smaller than
other systematic effects. So we have retained the original estimators $\RAP$,
$\Rm$ and $R_{Z}$ in the present work.

As previously stated, improvement coefficients are short distance quantities,
which can be determined in small physical volumes, using the Schr\"odinger
functional setup, with $L^3 \times T$ lattices having periodic boundary
conditions in space and Dirichlet boundary conditions in time. Definitions
of boundary operators and related correlation functions are given in
appendix~\ref{app:SF}. For reasons explained above, sea quark masses are tuned
closely to the chiral limit, in line with the usual ALPHA Collaboration choice
of mass-independent renormalisation schemes.

\subsection{The strategy revisited}
\label{subsec:newstrategy}

In refs.~\cite{ Guagnelli:2000jw,Fritzsch:2010aw} the computation of the
Symanzik $b$-coefficients proceeds as follows: the PCAC quark masses 
$\mij{11}$, $\mij{22}$, $\mij{12}$ and $\mij{33}$ are first determined in
standard fashion
from eq.~\eqref{eqn:mpcac_x0-ij}, and are subsequently fed into the ratios
$R_X$ (with $X={\rm AP}, {\rm m}, Z$) defined in eqs.~\eqref{eqn:estim-RAP},
\eqref{eqn:estim-Rm} and \eqref{eqn:estim-RZ}. 
In some cases, this procedure turns out to suffer from numerical instabilities:
the numerators of $R_X$ are current quark mass differences, i.e., constructed
so that the leading contributions in powers of the lattice spacing $a$ cancel,
isolating the $b$-counter-terms.
If these delicate cancellations in the mass differences occur with inadequate 
precision, the signal may be lost to the noise. 
Moreover, as we decrease the heavier quark mass $\mqn{2}$ towards
$\mqn{1}$, striving to reduce discretisation effects, both numerator and
denominator of the three estimators $R_X$ will contrive to give a noisy signal.
We will show in appendix~\ref{app:oldstrategy}
(cf.~figure~\ref{img:Rfrommassfit}) examples of this instability. In order to
overcome this problem, we introduce here a more elaborate method of
analysis of quark masses evaluated from correlator measurements,
which should ameliorate the stability of the results.

We start with some general considerations. At fixed gauge coupling, the bare
PCAC quark masses $m_{ij}$ defined in eq.~\eqref{eqn:PCAC-mass} depend on the 
subtracted valence quark masses $m_{{\rm q},i}$ and $m_{{\rm q},j}$ and the trace 
of the sea quark mass matrix
$\Tr{\Mq}$.  Since in our simulations sea quark masses are very close to the
chiral limit but not strictly zero, in what follows we keep $\Tr{\Mq}$
terms in the equations. The current masses $m_{ij}$ are symmetric functions
under the exchange $m_{{\rm q},i} \leftrightarrow m_{{\rm q},j}$. This implies
that they can be expressed as a power series of the form
\begin{align}\label{eq:mPCAC}
   am_{ij}( a\mqn{ij}, a\Delta_{ij} ) &= \sum_{n,k=0}^\infty  C_{nk} (a\Delta_{ij})^{2n}  (a\mqn{ij})^k  \;,
\end{align}
with the mass-splitting denoted as 
$\Delta_{ij}\equiv\tfrac{1}{2}(\mqn{i}-\mqn{j})$ and the dimensionless 
coefficients as $C_{nk}$. The latter only depend on the gauge
coupling and flavour-blind traces of the sea quark masses.

To next-to-leading order in the lattice spacing, the Symanzik
expansion for this expression is given by eq.~\eqref{eq:massmatch}.
A comparison with eq.~\eqref{eq:mPCAC} shows that, to this order, 
the expansion coefficients read
\begin{subequations}\label{eq:Cij}
\begin{align}\notag
C_{00}  &= Z \,\dfrac{a\Tr{\Mq}}{\NF} \Bigg \{ (r_{\rm m}-1) + (r_{\rm m} d_{\rm m}   - \bm) \dfrac{a\Tr{\Mq^2}}{\Tr{\Mq}}   \\\label{eq:C00}
&\qquad\qquad\qquad + \Big ( (r_{\rm m} \bar d_{\rm m} - \bar b_{\rm m}) - (r_{\rm m}-1) (\bar b_{\rm A} - \bar b_{\rm P}) \Big ) a\Tr{\Mq} \Bigg\}  \;, \\\label{eq:C01}
C_{01}  &= Z  \,\Big \{ 1+ \Big ( \bar b_{\rm m} -  (\bar b_{\rm A} - \bar b_{\rm P}) - (\ba-\bp) {(r_{\rm m}-1)}/{\Nf}   \Big )    a\Tr{\Mq} \Big \}  \;, \\\label{eq:C02}
C_{02}  &= Z  \,\Big \{ \bm - (\ba - \bp) \Big \}  \;,  \\\label{eq:C10}
C_{10}  &= Z  \,\bm    \;.
\end{align} 
\end{subequations}
If the sea quark masses were tuned {\it exactly} to their critical value, we
would have $C_{00} = 0$ and $C_{01} = Z$. Moreover, all $C_{nk}$ would be
functions of only $g_0^2$ in perturbation theory.
A non-perturbative determination of the $C_{nk}$, however, depends on the
imposed improvement condition.

Turning next to the specific case under study, we recall that the ratios
$R_{X}$ require three quark masses. We define the lightest one to
be $\mqn{1}$ and set it to the value of the three degenerate sea quark masses
used in our simulations, thus having $\mqn{1}=\Tr{\Mq}/\Nf$.  In practice, its
value is very small, but not strictly zero within statistical errors, cf.
the $x$-axis of figure~\ref{img:chiexp}. The heavier mass is $m_{{\rm q},2}$ 
and the average of the two is $m_{{\rm q},3}$, i.e., we have $m_{{\rm q},2} >
m_{{\rm q},3} > m_{{\rm q},1}$. Starting from eq.~(\ref{eq:mPCAC}), we now
write the current quark masses $m_{11}, m_{22}$ and $m_{12}$ as power series,
with $m_{22}$ and $m_{12}$ re-expressed in terms of $\mqn{1}$ and the
(partially-quenched) mass-splitting $\Delta \equiv \tfrac{1}{2}(\mqn{2} -
\mqn{1}) = \mqn{3} - \mqn{1}$:
\begin{align}\label{eq:m11}
am_{11}         &= \sum_{k=0}^\infty   C_{0k} (am_{\rm q, 1})^k \; ,                 \\\label{eq:m22}
am_{22}(\Delta) &= \sum_{k=0}^\infty   C_{0k} (am_{\rm q, 1} + 2a\Delta)^k \; ,      \\\label{eq:m12}
am_{12}(\Delta) &= \sum_{n,k=0}^\infty C_{nk} (a\Delta)^{2n} \big( a\mqn{1}+ a\Delta \big)^k \; .
\end{align}
We observe that their first derivatives are exactly related via
\begin{align}\label{eq:deriv-m}
        \frac{1}{2}\frac{\partial m_{22}}{\partial \Delta}\bigg|_{\Delta=0} &= \frac{\partial m_{12}}{\partial \Delta}\bigg|_{\Delta=0} = \frac{\partial m_{11}}{\partial \mqn{1}}    \;.
\end{align}%
By construction, the unitary setup is recovered when $\Delta\to 0$, in which 
case ${m_{12},m_{22}\to m_{11}}$.
Written in this way, we can always expand the masses $am_{12}$ and $am_{22}$ 
close to $am_{11}$, where then $\Delta$ becomes the expansion parameter:
\begin{align}\label{eq:power-m12}
    am_{12}(\Delta) &= am_{11}  +\hphantom{2} N_1 a\Delta  +\hphantom{4} N_2 (a\Delta)^2  +  \Or(\Delta^3)   \; ,  \\\label{eq:power-m22}
    am_{22}(\Delta) &= am_{11}  +          2  N_1 a\Delta  +          4  D_2 (a\Delta)^2  +  \Or(\Delta^3)   \; .
\end{align}
The coefficients $N_i$ and $D_i$ for the flavour non-diagonal and diagonal 
masses, respectively, are linear combinations of the $C_{nk}$ and carry a 
residual dependence on $a\mqn{1}$.  Our par\-ti\-cular choice of the third mass,
$\mqn{3}\equiv \mqn{12}$, leads to the identity 
$m_{33}(\Delta)\equiv m_{22}(\Delta/2)$. Accordingly, we have the expansion 
\begin{align}\label{eq:power-m33}
    am_{33}(\Delta) &= am_{11}  + N_1 a\Delta  +  D_2(a\Delta)^2  +  \Or(\Delta^3)   
\end{align}
at our disposal and can revisit eqs.~\eqref{eqn:estim-RX} in the context of our
current discussion: 
\begin{subequations}\label{eq:RX2}
\begin{align}\label{eq:RAP2}
  R_{\rm AP} &\equiv  \dfrac{2 m_{12} - m_{11} - m_{22}}{(m_{22} - m_{11})\,a\Delta} 
              = \dfrac{ N_2 - 2D_2 + \Or(a\Delta)}{N_1+ \Or(a\Delta)}  
              &\xrightarrow{\Delta,\mqn{1} \to 0} &\qquad \ba-\bp  \;, \\\label{eq:Rm2}
         \Rm &\equiv  \dfrac{2\left(\mij{12}-\mij{33}\right)}
              {\left(\mij{22}-\mij{11}\right)\,a\Delta}   =  \dfrac{ N_2 - D_2+ \Or(a\Delta)}{ N_1+ \Or(a\Delta)} 
              &\xrightarrow{\Delta,\mqn{1} \to 0} &\qquad \bm \;, \\
          R_Z &\equiv  \dfrac{\mij{22}-\mij{11}}{2\Delta} + [\RAP-\Rm](a\mij{11}+a\mij{22}) \notag \\\label{eq:RZ2}
                     &\qquad\qquad\qquad =  N_1 - \frac{2 D_2 am_{11}}{{N_1}}  +  \Or\big((a\Delta)^2\big)  %\\
                     &\xrightarrow{\Delta,\mqn{1} \to 0} &\qquad Z \;. 
\end{align}
\end{subequations}

The original estimator $R_Z$ was constructed in order to cancel an
$\Or(a\Delta)$ effect~\cite{Guagnelli:2000jw}, i.e., to reduce the largest bias
in the determination of the leading order factor $Z$. In the unquenched theory,
uncancelled $\Or(a\Tr{\Mq})$ effects remain in all quantities. They are
typically supressed by 1\,--\,2 orders in magnitude due to the sea quark mass
tuning ($m_{11}\approx 0$) required in a mass-independent 
renormalisation scheme. In the determination of the estimators $R_X$, 
a renormalised trajectory, or line of constant physics (LCP), has to be 
employed such that they adopt the proper scaling behaviour when the bare gauge 
coupling $g_0^2$ is varied.
This means that, besides $m_{11}=0$, a value for the mass-splitting
$\Delta$, which fixes the LCP in the valence sector, has to be specified.
In principle, any sensible choice $\Delta\ne 0$ is sufficient to define
a valid set $\{R_{\rm AP},R_{\rm m},R_{Z}\}_\Delta$ that achieves 
$\Or(a)$ improvement in physical quantities. 
Different choices lead to somewhat different approaches to the continuum limit
and are equivalent in the framework of Symanzik's effective theory. Their
relative difference is a higher-order cutoff effect that vanishes for $a\to 0$
as can be easily seen in eqs.~\eqref{eq:RX2}. A non-perturbative determination
of these estimators inherits a non-trivial all-order dependence on $a\Delta$ if
the limit $\Delta\to 0$ is not taken. In that sense, an explicit choice of
$\Delta$ constitutes an ambiguity in their definition. In
ref.~\cite{Fritzsch:2010aw}, for instance, $\Delta$ has been held constant by
requiring $Lm_{22}\approx 0.5$ at constant physical $L$ with $L/a\in[12,24]$.

In the present paper we aim at eliminating this $\Delta$-ambiguity in the
definition of $R_{\rm AP}$, $R_{\rm m}$ and $R_{Z}$, because it can
potentially lead to larger cutoff effects in the physics of light quarks.  By
noting that eqs.~\eqref{eq:power-m12} and~\eqref{eq:power-m22} for $am_{ij}$ 
can literally be used as joint ansatz for interpolating fit functions 
(polynomials in $a\Delta$), we are able to build the standard estimators 
according to eqs.~\eqref{eq:RX2} by dropping the $\Or(a\Delta)$ terms 
explicitly. 
In this case only the first few parameters of the polynomials are relevant,
which can be well controlled by sufficiently scanning the diagonal and 
non-diagonal current quark masses as functions of $a\Delta$. 
The sub-leading effects in the sea quark mass of order $a\Tr{\Mq}$ can be 
removed by extrapolation or interpolation.
The presented proposal has the additional advantage that no iterative tuning 
of the second (and subsequently third) mass is required in advance. 

We will refer to the results of this analysis as LCP-0, since they
are obtained along the line of constant physics which keeps all the masses 
equal to zero:
\begin{align}\label{eq:LCP0}
        \text{LCP-0:} \qquad 
        L=\text{const}\;, \quad Lm_{11}=0\;, \quad L\Delta_{22} =0 \;.
\end{align}
Here we have introduced the current quark mass difference
\begin{align}\label{eq:Delta22}
        L\Delta_{22} &\equiv L(m_{22}(\Delta)-m_{11}) \;,
\end{align}
which is in one-to-one correspondence with the difference of bare subtracted 
quark masses $a\Delta$ and reduces to $Lm_{22}$ in the chiral limit, $m_{11}=0$.

Besides the determinations in the massless unitary setup, we will also give 
results for massive valence quarks. 
In fact, experience shows that in large-volume simulations, heavy-flavour
Wilson quarks have sizeable mass-dependent cutoff effects in the
typical range of lattice spacings $0.04 \lesssim a/\fm \lesssim 0.1$.
For that reason one may favour the opposite interpretation and
exploit the freedom in Symanzik's effective theory to determine the improvement
functions $R_X$ at a value of $\Delta$ that is as close as possible to the 
characteristic heavy quark mass scale typically involved in the application in
question.  
By doing so, the interpolating functions for the PCAC
masses, eqs.~\eqref{eq:power-m12}--\eqref{eq:power-m33}, have to be evaluated 
at $\Delta\ne0$ and fed into the defining expressions for the estimators.
At the non-perturbative level, this corresponds to a resummation of all 
higher-order terms in $a\Delta$ for the chosen line of constant physics. 
The effectiveness of this approach was demonstrated in 
ref.~\cite{Fritzsch:2010aw}, where two determinations of $R_X$ at 
$Lm_{22}\approx 0.5$ and $Lm_{22}\approx 2.5$ were
probed in the heavy quark sector with masses above and below the bottom quark
mass, finding  a more significant reduction of mass-dependent cutoff effects 
and an extension of the $a^2$-scaling region in the case of the
largest~$\Delta$.
Therefore, we also introduce a second line of constants physics in the
valence sector,
\begin{align}\label{eq:LCP1}
        \text{LCP-1:} \qquad 
        L=\text{const}\;, \quad Lm_{11}=0\;, \quad L\Delta_{22}=1 \;.
\end{align}
Thus we will determine a second set of estimators $R_X$, suitable for 
calculations with $2+1$ dynamical light quarks and valence charm quarks.

In sect.~\ref{sec:analysis} we will elaborate on both variants of the
data analysis and the achieved control over the systematic effects.

\section{Gauge configuration ensembles}
\label{sec:ensembles}

\begin{table}[t]
    \centering\small
	\renewcommand{\arraystretch}{1.25}
    \begin{tabular}{lccllrrrll}
\toprule
        ID & $\textstyle{\frac{L}{a}}$ & $\textstyle{\frac{T}{a}}$ & $\beta$ & $\kappa_1$ & $N_{\rm r}$ & $\Ncfg$ & $\Ncfg^{(0)}$ & $am_{11}$ & $am_{11}^{(0)}$\\
\midrule
A1k1 & $12$ & $17$ & $3.3$ & $0.13652$ & $20$ & $2560$ & $935$ & $-0.00166(61)$ & $-0.00278(80)$\\
A1k3 & $12$ & $17$ & $3.3$ & $0.13648$ & $5$ & $1719$ & $614$ & $\phantom{+}0.00262(130)$ & $\phantom{+}0.00079(118)$\\
A1k4 & $12$ & $17$ & $3.3$ & $0.13650$ & $20$ & $12080$ & $4424$ & $\phantom{+}0.00030(29)$ & $-0.00110(36)$\\
\hline
E1k1 & $14$ & $21$ & $3.414$ & $0.13690$ & $32$ & $4800$ & $1694$ & $\phantom{+}0.00308(22)$ & $\phantom{+}0.00262(26)$\\
E1k2 & $14$ & $21$ & $3.414$ & $0.13695$ & $47$ & $7050$ & $2653$ & $\phantom{+}0.00034(18)$ & $-0.00022(22)$\\
\hline
B1k1 & $16$ & $23$ & $3.512$ & $0.13700$ & $3$ & $3328$ & $1336$ & $\phantom{+}0.00562(14)$ & $\phantom{+}0.00549(21)$\\
B1k2 & $16$ & $23$ & $3.512$ & $0.13703$ & $2$ & $1151$ & $395$ & $\phantom{+}0.00481(19)$ & $\phantom{+}0.00444(25)$\\
B1k3 & $16$ & $23$ & $3.512$ & $0.13710$ & $2$ & $2048$ & $938$ & $\phantom{+}0.00164(16)$ & $\phantom{+}0.00107(20)$\\
B1k4 & $16$ & $23$ & $3.512$ & $0.13714$ & $1$ & $3482$ & $1401$ & $\phantom{+}0.00002(14)$ & $-0.00057(19)$\\
\hline
C1k2 & $20$ & $29$ & $3.676$ & $0.13700$ & $4$ & $1904$ & $857$ & $\phantom{+}0.00619(7)$ & $\phantom{+}0.00600(11)$\\
C1k3 & $20$ & $29$ & $3.676$ & $0.13719$ & $4$ & $1934$ & $1249$ & $-0.00086(8)$ & $-0.00109(11)$\\
\hline
D1k2 & $24$ & $35$ & $3.810$ & $0.13701$ & $2$ & $803$ & $357$ & $\phantom{+}0.00084(8)$ & $\phantom{+}0.00079(10)$\\
D1k4 & $24$ & $35$ & $3.810$ & $0.137033$ & $8$ & $5313$ & $3469$ & $-0.00002(3)$ & $-0.00007(3)$ \\ \bottomrule
\end{tabular}

    \caption{Overview of the simulation parameters of the $\Nf=3$ ensembles
        (labeled by ID) that represent our data. Subsequent columns refer to
        the lattice dimensions $L^3T/a^4$, the inverse gauge coupling
        $\beta=6/g_0^2$, the light (sea) quark hopping parameter~$\kappa_1$, 
        the number of replica $N_{\rm r}$, 
        the number of configurations per ensemble, both in total ($\Ncfg$)
        and in the subset of configurations with zero topological
        charge ($\Ncfg^{(0)}$), and the corresponding PCAC sea quark masses.
        All ensembles have configurations separated by 8 molecular dynamic
        units (MDU), except for A1k3 and D1k4 that have 4 and 16 MDU,
        respectively.
        Compared to the data base of \cite{Bulava:2015bxa, Bulava:2016ktf}, we
        have generated and used the nearly chiral ensembles A1k3, A1k4, B1k4
        and D1k4, and significantly increased statistics for E1k1 and E1k2.
       }
    \label{tab:enstab}
\end{table}

The three-flavour lattice QCD simulations in the Schr\"odinger functional
framework have been performed using the \texttt{openQCD} code of 
ref.~\cite{openQCD}, with tree-level Symanzik-improved gauge
action~\cite{Luscher:1984xn}, $\Nf=3$ massless Wilson-clover fermions,
vanishing boundary gauge fields $C=C'=0$ and boundary fermion parameter
$\theta=0$. The value of the improvement coefficient $\csw$ is taken from
ref.~\cite{Bulava:2013cta}. The RHMC
algorithm~\cite{Kennedy:1998cu,Clark:2006fx,Luscher:2008tw} is used for the
third dynamical quark. The relevant modification of the integration measure of
the fermion determinant is then compensated by the inclusion of a reweighting
factor in the analysis.

Most of the ensembles in this study coincide with those of
refs.~\cite{Bulava:2015bxa,Bulava:2016ktf}, where the improvement coefficient
$\ca$ and the normalisation constant $\za$ of the axial vector current are
determined. In these works the constant physics condition is fixed by setting
${L \approx 1.2\,\fm}$. This is achieved by beginning with a particular pair of
$g^2_0$ and $L/a$ ($\beta= 6/g^2_0=3.3$ at $L/a =12$ here). 
Next we chose the bare couplings for subsequent smaller lattice spacings 
according to the universal two-loop $\beta$-function. 
In this way, lattice spacings are covered
in the range from $a \approx 0.09\,\fm$ to $a \approx 0.045\,\fm$. At each bare
coupling, we generate ensembles for a few small values of the bare sea current
quark mass $am_{11}$, in order to obtain an estimate of the critical point
$\kappa_{\rm crit}$ and to be able to extrapolate to $am_{11}=0$ at a subsequent 
stage of the analysis.

Table~\ref{tab:enstab} gives an overview of the ensembles used in this work. 
The labelling of these ensembles, based on an alphanumeric four-symbol code
such as A1k1, follows the conventions: the first letter (A-E) represents
a specific lattice geometry $L^3 T/a^4$, while diffe\-rent choices of $\beta$ 
for a given geometry are distinguished by the subsequent number. In the present
work, we have a single $\beta$ for each geometry. Separated by a ``k'', the
final integer labels the sea quark hopping parameter $\kappa_1 = \kappasea$.
In addition to the ensembles available to us from previous ALPHA Collaboration
simulations~\cite{Bulava:2015bxa,Bulava:2016ktf}, we have generated ensembles
A1k3, A1k4, B1k4, D1k2 and D1k4, with $\kappasea$ tuned so that the
corresponding PCAC masses are closer to the chiral limit. Furthermore, the
replica lengths of ensembles E1k1 and E1k2 were increased for larger statistics
and a more reliable estimation of autocorrelations.

Note that the values of $\beta$ are in the same range as those of the
large-volume ensembles produced with the same lattice action by the
CLS (Coordinated Lattice Simulations)
effort~\cite{Bruno:2014jqa,Bruno:2016plf,Bali:2016umi}.
Therefore, our results can be applied in, e.g., determinations of $\Or(a)$
improved phenomenological quantities such as quark masses and decay constants.

\subsection{Topological charge}

\begin{figure}[]
	\includegraphics[width=\linewidth]{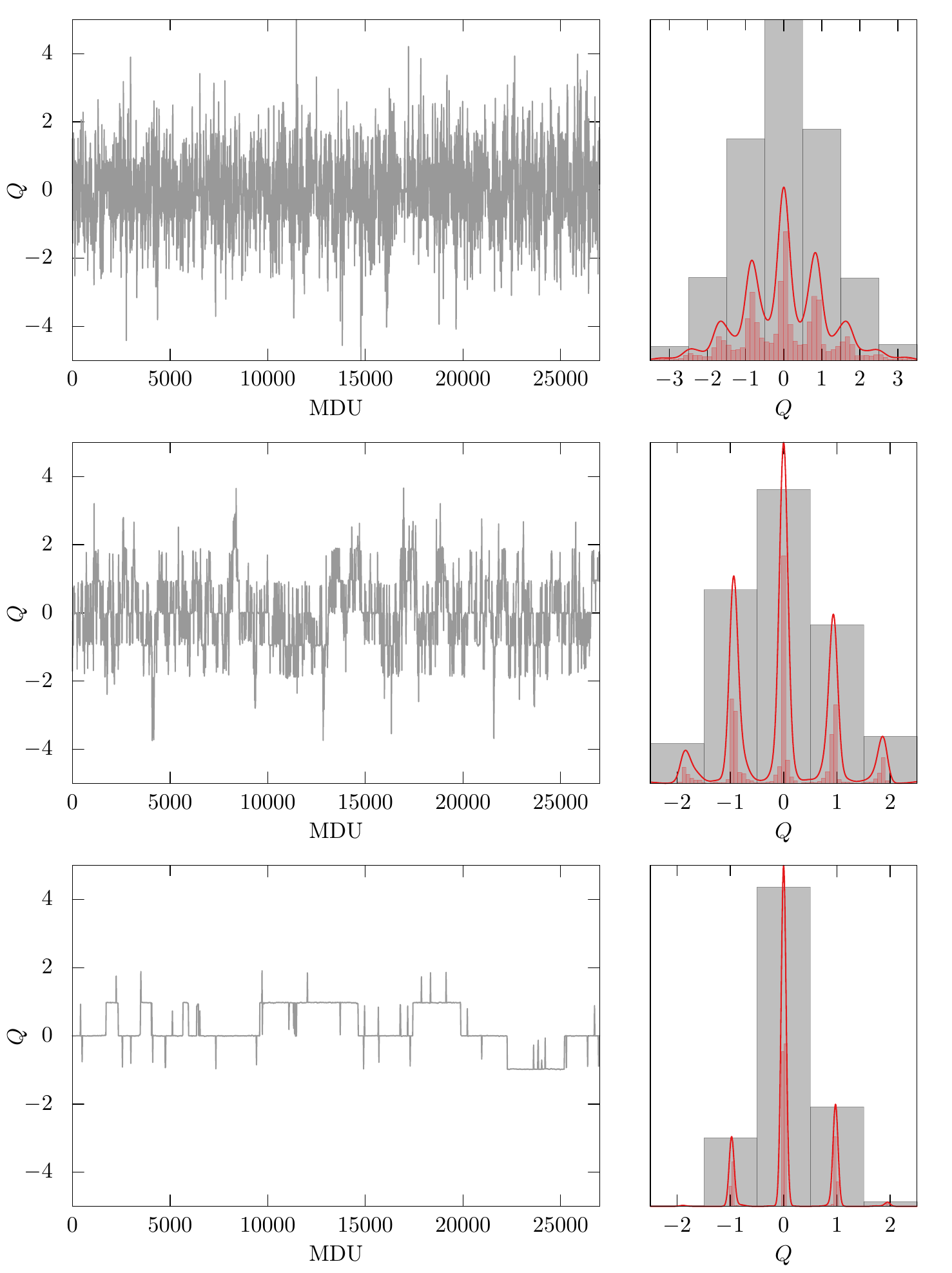}
    \caption{Topology freezing, monitored through Monte Carlo histories and 
        distributions of the topological charge $Q$ for decreasing lattice 
        spacing (top to bottom: ensembles A1k4, B1k4
        and D1k4). The grey histogram counts appearances of $Q$ beloging to
        different integer sectors $\nu\in\mathbb{Z}$ according to
        $\delta_{Q-\nu,0}$ of eq.~\eqref{eq:qtop}. The finer-spaced (red) 
        histograms reveal the fine-structure of the (non-integer) topological
        charge distribution with Wilson fermions and gradient-flow smoothing
        ratio $c=0.35$. Red curves are naive fits to a sum of Gaussian 
        distributions.
		}\label{img:qtop}
\end{figure}

In QCD with Schr\"{o}dinger functional boundary conditions, disconnected
topological sectors emerge in the continuum limit. However, for small or
intermediate physical volumes as employed here, non-trivial topological
sectors, i.e., those with topological charge ${Q\neq 0}$, only receive a small
weight in the partition sum.
This issue of topology freezing, previously investigated in
refs.~\cite{Fritzsch:2013yxa,Bulava:2015bxa,Bulava:2016ktf,DallaBrida:2016kgh},
may be met by projecting the quantities of interest to the $Q=0$ topological 
sector.
For the case at hand, involving improvement coefficients and renormalisation 
constants, quantities defined in this way differ from their full-topology 
counterparts only by irrelevant cutoff effects and exhibit a smooth approach 
to the continuum limit.

Figure~\ref{img:qtop} shows Monte Carlo histories of the topological charge $Q$
and its distribution on three exemplary ensembles. 
The effect of topology freezing is clearly visible:
while the topological charge is appropriately sampled for the coarse lattice
spacing in the A1 ensembles (top), the HMC algorithm is not able to properly
tunnel between different topological charge sectors at finer lattice spacings
($L\approx\mathrm{const}$); this becomes most pronounced for the D-ensembles 
(bottom).
Sectors with $Q=-1,0,1$ are mostly sampled, and the charge remains for a 
longer Monte Carlo time in single sectors when the lattice spacing is 
decreased.
Such a behaviour is in line with similar findings, e.g.,
in \cite{Bulava:2015bxa,Bulava:2016ktf,DallaBrida:2018tpn}.
As in these references, we thus have confined the analysis to the sector with
zero topological charge, in order to avoid potential bias from improper
sampling and associated, unresolved large autocorrelation times that could
affect a reliable statistical error estimation for our observables.
Note that this procedure is also theoretically sound, since our strategy to
extract $\bm$, $\ba-\bp$ and $Z$ relies on PCAC quark masses defined through
Ward identities which, being operator relations, hold also within a single 
topological sector.
Although this projection to zero topological charge typically comes at the
expense of larger statistical uncertainties and a slightly modified cutoff 
dependence, it is not expected to induce a noticeable difference in the final 
results.
This is indeed confirmed in~tables~\ref{tab:restab} and \ref{tab:restab_Lx1}.

The projection of an observable $O$ onto the sector of trivial ($Q=0$)
topology was introduced in \cite{Fritzsch:2013yxa,DallaBrida:2016kgh}
via
\begin{align}\label{eq:qtop}
        \langle O\rangle_0 &= \frac{\langle O \delta_{Q,0} \rangle}{\langle\delta_{Q,0}\rangle} \;, &
        \delta_{Q,0} &\rightarrow \Theta(Q+0.5)\Theta(0.5-Q) \;, 
\end{align}
where $\delta_{Q,0}$ is replaced by step functions,
because the topological charge takes non-integer values in finite volume.
We adopt the charge defined via the gradient flow as in
ref.~\cite{Luscher:2010iy}, at gradient flow time corresponding to a
smoothing ratio of $c\equiv\sqrt{8t}/L = 0.35$. 
For comparison, we also quote the results for the analysis including all
topological sectors.

Further details pertinent to our ensembles can be found in 
refs.~\cite{Bulava:2015bxa,Bulava:2016ktf}, which report ${\nf=3}$
calculations of $\ca$ and $\za$. These concern the implementation of the
line of constant physics, the negligible influence of its small violations on
the results, the simulation algorithm used to generate the gauge
configurations, and the projection onto the trivial-topology sector.

\subsection{Statistical error analysis}

The statistical uncertainties are determined using the $\Gamma$-method
\cite{Wolff:2003sm, Schaefer:2010hu}, so as to take the autocorrelations of
all observables into account. An independent analysis, using jackknife error 
estimation, was done as follows. First the replica of an
ensemble are concatenated and subsequently subdivided into bins of width ten.
Then, the standard jackknife error is computed by eliminating a single bin
average at a time. The bin width was specified by varying the bin size and
choosing the minimal value at which the jackknife error stabilises.
Error estimates from both methods are in very good agreement.

The chiral extrapolations discussed in the next section are based on
independent datasets of ensembles belonging
to the same group (e.g. of A1k1, A1k3 and A1k4 belonging to the A1-group).
In the context of the jackknife error analysis, we exploit the embedding trick
for combining statistically independent runs, described in appendix A.3 of
\cite{DelDebbio:2007pz}.

\section{Data analysis}
\label{sec:analysis}

The definitions of the Schr\"odinger functional (SF) correlation functions 
$\fa$, $\fp$ and the PCAC quark masses $m_{ij}$ are standard; therefore,
we defer all details to appendix~\ref{app:SF}.

For each ensemble we compute valence quark propagators for $\mqn{1}$ 
(which is fixed by the sea quark hopping parameter of the simulation)
and for $\Or(15)$ values of $\mqn{2}$ in the range ${0\leq L\Delta_{22}\leq 1}$,
as well as for the corresponding $\mqn{3}\equiv\tfrac{1}{2}(\mqn{1}+\mqn{2})$. 
Earlier approaches~\cite{Guagnelli:2000jw,Fritzsch:2010aw} relied upon the
three distinct masses, in order to evaluate
the estimators $R_X$ by direct use of eqs.~(\ref{eqn:estim-RX}).
In the strategy adopted in the
present work, $m_{33}$ is simply another current quark mass diagonal in flavour,
so it is on an equal footing with $m_{22}$, thereby enriching the density of
points in the low mass region. Results from the earlier method and the issues
related to it are discussed in appendix~\ref{app:oldstrategy}.

The SF correlation functions for the appropriate mass combinations are 
obtained be utilizing the \texttt{sfcf} program \cite{master:Wittemeier}.

We compute the PCAC masses $m_{11}$, $m_{22}$, $m_{12}$ and $m_{33}$ 
for each time-slice $x_0$ using the improved lattice derivatives of 
eq.~(\ref{eqn:deriv}). 
Then we average over the middle third of the time extent 
$T=(3/2)L$, i.e., $x_0/a\in [L/(2a),L/a]$ (keeping the physical plateau 
length constant); an example is shown in figure~\ref{img:B1k4_plateau}.
The more standard choice of the mass definition at $x_0=T/2$ with standard
derivatives has been taken into account for comparison.

\subsection{Interpolating functions for PCAC quark masses}
\label{subsec:massinterpolations}

\begin{figure}[t]
    \includegraphics[width=\linewidth]{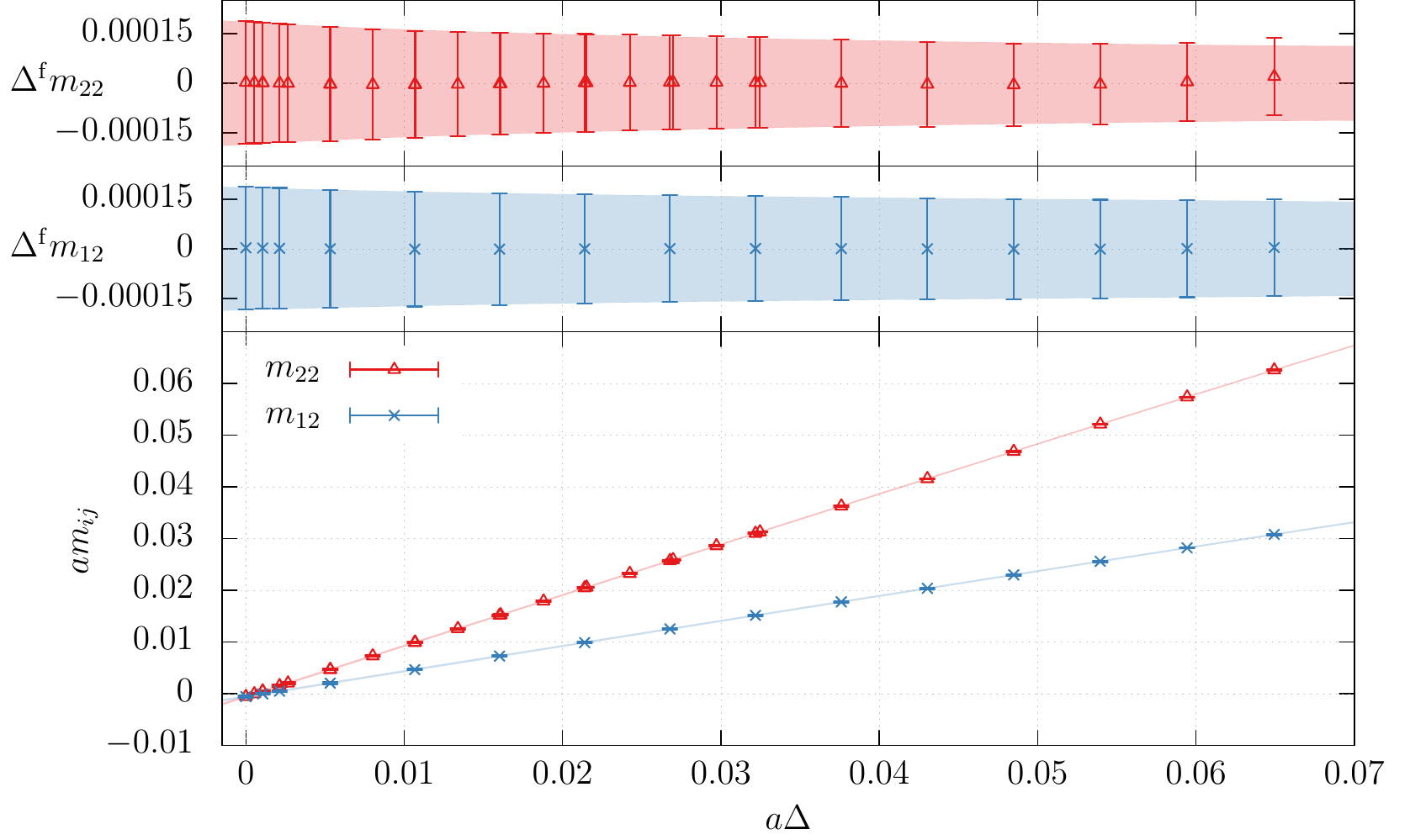}
    \caption{\textit{Lower plot:} Example of a combined mass fit (ensemble
        B1k4) depicting the fitted data points and curves for $m_{22}$ and
        $m_{12}$. Error bars and bands are too small compared to the scale of
        the plot. The density of $m_{22}$ points increases at smaller $m_{22}$,
        due to the inclusion of $m_{33}$ results.
        \textit{Upper plots:} Differences between the measured PCAC masses
        $m_{ij}$ and the fitted curves, ${\Delta^{\rm f} m_{ij} =
        m_{ij}(\kappa_2)-m_{ij}(\Delta)}$ together with the statistical
        uncertainties of the data points and the uncertainty band of the
        curves.
    }
    \label{img:massfit}
\end{figure}

We proceed by applying the method described in 
subsect.~\ref{subsec:newstrategy}.  
With the measurements done as explained above, we have $\Or(15)$ estimates of
flavour non-diagonal masses $m_{12}$ and $\Or(30)$ estimates of diagonal ones
$m_{22}$ and $m_{33}$.
An example is shown in figure~\ref{img:massfit}, where data points for 
diagonal and non-diagonal PCAC masses are plotted as functions of
${\Delta\equiv\tfrac{1}{2}(\mqn{2}-\mqn{1})}$ for the nearly chiral
ensemble B1k4.
To simplify the notation, all quantities connected with and derived from
the diagonal masses $m_{22}$ and $m_{33}$ will henceforth be denoted by
the subscript ``22''.

Following eqs.~\eqref{eq:power-m12} and~\eqref{eq:power-m22}, we minimise for
the parameters of two polynomials of a given degree ($am_{11}$, $N_1$, $N_2$,
$\ldots$, $D_2$, $\ldots$), constrained to have the same intercept and
related first derivatives. In order to avoid over-constraining our fits,
we prefer treating $am_{11}$ as a free parameter, rather then keeping it fixed
to its measured mean value.

We have opted for third-order polynomials. 
From eqs.~(\ref{eq:RAP2})--(\ref{eq:RZ2}) we see that for the determination
of the estimators at the unitary point (i.e., $a\Delta=0$) we only need the
polynomial coefficients up to second order.
By fitting with third-order polynomials we take into account possible
higher-order effects, without contaminating the lower order coefficients.
The dependence of the results on the polynomial order is investigated in
subsect.~\ref{subsec:ambigpolydegree}.

In the two upper panels of figure~\ref{img:massfit} we 
display the difference between the interpola\-ting fit and the data
points for the diagonal and non-diagonal masses, respectively.
Comparing the deviation of the mean values from zero, we see that the combined
fit is an excellent representation of the data down to $10^{-5}$.

\subsection{Estimators from the interpolation method}
\label{subsec:estimators}

\begin{figure}[t] 
    \centering
    \includegraphics[width=.75\linewidth]{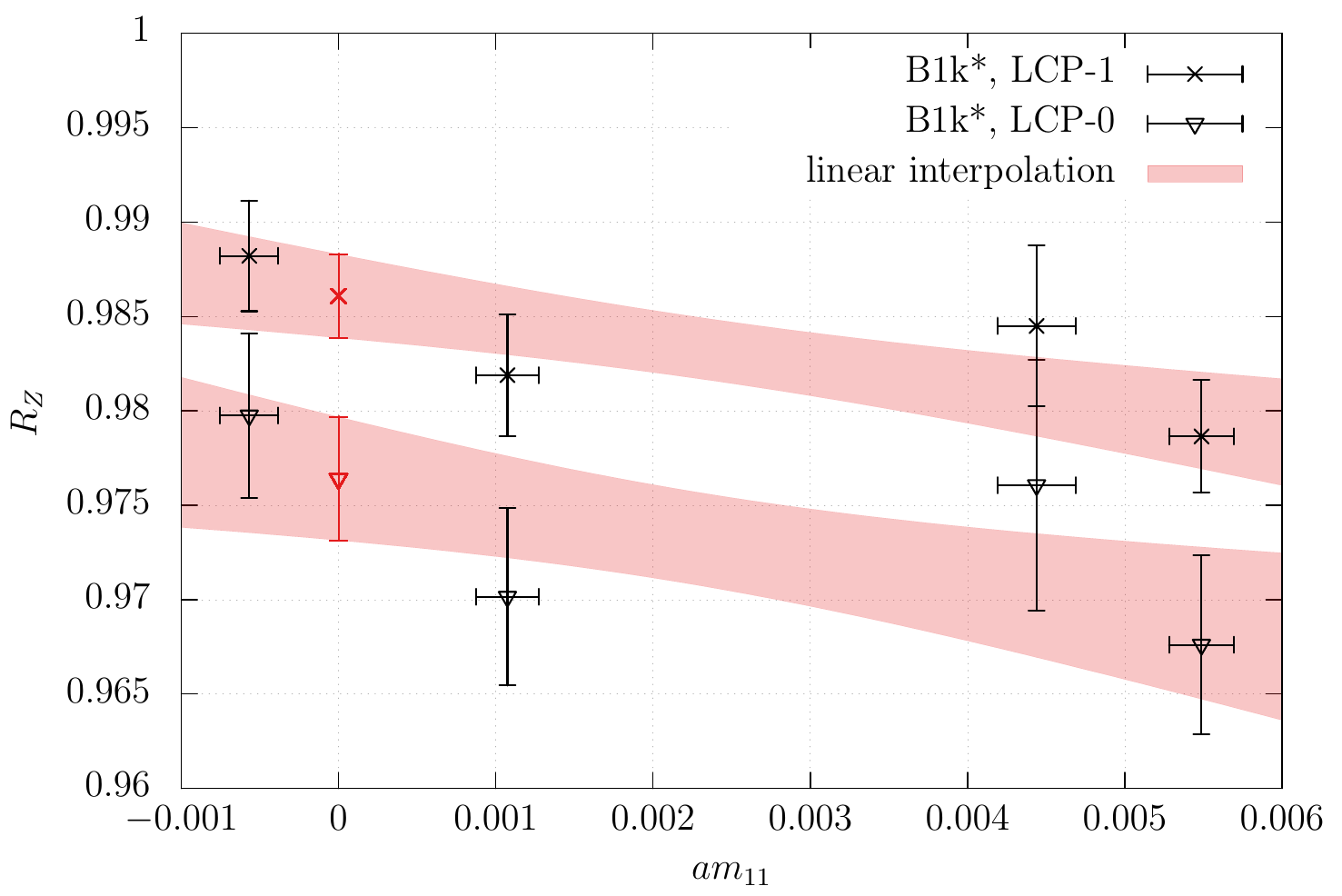}
    \caption{Chiral interpolation ($am_{11}\to 0$) of $R_Z$ for $\beta=3.512$
    for the two lines of constant physics defined in
    eqs.~(\ref{eq:LCP0}) and (\ref{eq:LCP1}).
    The red points represent the resulting values in the chiral limit.
    }
    \label{img:chiexp}
\end{figure}
\begin{table}[t]%
    \centering\small
    \renewcommand{\arraystretch}{1.25}
    \begin{tabular}{cllllll}
\toprule
$\beta$ & $R_\text{AP}^{(0)}$ & $R_\text{AP}^{(\text{all})}$ & $R_\text{m}^{(0)}$ & $R_\text{m}^{(\text{all})}$ & $R_Z^{(0)}$ & $R_Z^{(\text{all})}$\\
\midrule
$3.300$ & $-0.769(101)$ & $-0.656(55)$ & $\phantom{+}1.303(90)$ & $\phantom{+}1.244(43)$ & $\phantom{+}0.7462(56)$ & $\phantom{+}0.7468(28)$\\
$3.414$ & $-0.812(53)$ & $-0.770(53)$ & $\phantom{+}0.291(53)$ & $\phantom{+}0.364(44)$ & $\phantom{+}0.8762(40)$ & $\phantom{+}0.8719(37)$\\
$3.512$ & $-0.515(49)$ & $-0.536(36)$ & $-0.291(39)$ & $-0.177(32)$ & $\phantom{+}0.9764(33)$ & $\phantom{+}0.9672(26)$\\
$3.676$ & $-0.291(46)$ & $-0.279(37)$ & $-0.671(43)$ & $-0.583(35)$ & $\phantom{+}1.0588(31)$ & $\phantom{+}1.0536(23)$\\
$3.810$ & $-0.156(20)$ & $-0.144(17)$ & $-0.738(19)$ & $-0.700(18)$ & $\phantom{+}1.0882(11)$ & $\phantom{+}1.0866(10)$\\
\bottomrule
\end{tabular}

    \caption{Chirally extrapolated LCP-0 results, both
             for the vanishing topological charge sector, $\smash{R_X^{(0)}}$,
             and without zero-charge projection, $\smash{R_X^{(\text{all})}}$.
             Only the former are plotted in figure~\ref{fig:g0sq-results}.
    }
    \label{tab:restab}
\end{table}

Having determined the fit polynomials of non-diagonal and diagonal PCAC
masses, we evaluate the estimators $R_X$ at the two lines of constant physics,
LCP-0 and LCP-1, speci\-fied by eqs.~(\ref{eq:LCP0}) and (\ref{eq:LCP1}).

For the unitary case LCP-0 with $a\Delta=0$
(or, equivalently, $L\Delta_{22}=0$), these estimators are built from the  
parameters $N_1$, $N_2$ and $D_2$ according to eqs.~\eqref{eq:RX2}.
We gather these results for all ensembles 
in tables \ref{tab:resAllEns_Q0} and \ref{tab:resAllEns_Qall} of
appendix~\ref{app:results}.

The next step towards a fully massless calculation is to extrapolate the
results to the chiral limit $m_{11}=0$; note that the knowledge of the exact
value of the critical hopping parameter is not required here. 
Since an ensemble with a small negative $m_{11}$ exists for all gauge couplings
in the trivial topological sector with $Q=0$ (cf.~table~\ref{tab:enstab}),
this leads in practice to chiral \emph{inter}polations rather than
extrapolations.
The coefficients $N_i$ and $D_i$ of the expansions~\eqref{eq:power-m12}
and~\eqref{eq:power-m22} have an implicit dependence on $am_{11}$, which only 
affects $\rmO(a^2)$~terms of these expansions. 
Thus a linear fit appears to be sufficient for our purposes.
An example is reproduced for ensemble group B1 ($\beta=3.512$) in
figure~\ref{img:chiexp}, where chiral interpolations of $R_Z$ are
presented.%
\footnote{%
Alternatively, such an interpolation can first be performed on the individual
fit parameters, before combining them into $R_X$. This leads to exactly the
same results in the chiral limit.
}
Results are listed in table~\ref{tab:restab}, for all topological sectors 
and after projecting to $Q=0$.

We repeat the full analysis for LCP-1, i.e., $L\Delta_{22}=1$.
The errors of the respective estimators are typically smaller than in the
unitary case LCP-0, cf.~tables~\ref{tab:resAllEns_Q0_Lx1}
and~\ref{tab:resAllEns_Qall_Lx1} of appendix~\ref{app:results}.
The interpolation to vanishing sea quark mass $m_{11}=0$
is also illustrated in figure~\ref{img:chiexp} for $R_Z$ of the B1 ensembles,
where one can infer that the difference of the two chiral (red) points is
statistically significant and represents an $a\Delta$-ambiguity.
The results for all chiral estimators of LCP-1 are given in
table~\ref{tab:restab_Lx1}. 
\begin{table}%
	\centering\small
	\renewcommand{\arraystretch}{1.25}
	\begin{tabular}{cllllll}
\toprule
$\beta$ & $R_\text{AP}^{(0)}$ & $R_\text{AP}^{(\text{all})}$ & $R_\text{m}^{(0)}$ & $R_\text{m}^{(\text{all})}$ & $R_Z^{(0)}$ & $R_Z^{(\text{all})}$\\
\midrule
$3.300$ & $-0.356(24)$ & $-0.376(10)$ & $-0.025(19)$ & $-0.002( 7)$ & $\phantom{+}0.7896(36)$ & $\phantom{+}0.7846(16)$\\
$3.414$ & $-0.362(13)$ & $-0.363(12)$ & $-0.264(12)$ & $-0.237(10)$ & $\phantom{+}0.8992(26)$ & $\phantom{+}0.8950(24)$\\
$3.512$ & $-0.227(12)$ & $-0.244( 9)$ & $-0.469(11)$ & $-0.429( 9)$ & $\phantom{+}0.9861(23)$ & $\phantom{+}0.9785(18)$\\
$3.676$ & $-0.125(14)$ & $-0.133(12)$ & $-0.643(14)$ & $-0.607(12)$ & $\phantom{+}1.0611(23)$ & $\phantom{+}1.0564(17)$\\
$3.810$ & $-0.070( 7)$ & $-0.071( 6)$ & $-0.684( 7)$ & $-0.669( 6)$ & $\phantom{+}1.0884( 8)$ & $\phantom{+}1.0871( 8)$\\
\bottomrule
\end{tabular}

    \caption{Chirally extrapolated LCP-1 results, both
             for the vanishing topological charge sector, $\smash{R_X^{(0)}}$,
             and without zero-charge projection, $\smash{R_X^{(\text{all})}}$.
             Only the former are plotted in figure~\ref{fig:g0sq-results}.
	}
	\label{tab:restab_Lx1}
\end{table}

\subsection{Ambiguity checks}
\label{subsec:ambiguitychecks}

Before presenting the final results, we discuss the ambiguities 
arising from our specific choices of improvement conditions.
These consist in: the projection to
topological sectors, the exact definition of the current quark masses, and the
interpolating functions that relate them to the mass difference $a\Delta$.
All these choices are formulated in a way that respects the
constant physics condition among different ensembles.
They are part of the non-perturbative operational definitions of the $R_X$,
which influence the numerical values of our final results.
We will present below some representative examples of these systematic 
effects. They are found similar in size to those previously observed in the 
quenched~\cite{Guagnelli:2000jw} and in the two-flavour~\cite{Fritzsch:2010aw}
determinations.

\begin{figure}[t]
	\centering
    \includegraphics[width=\linewidth]{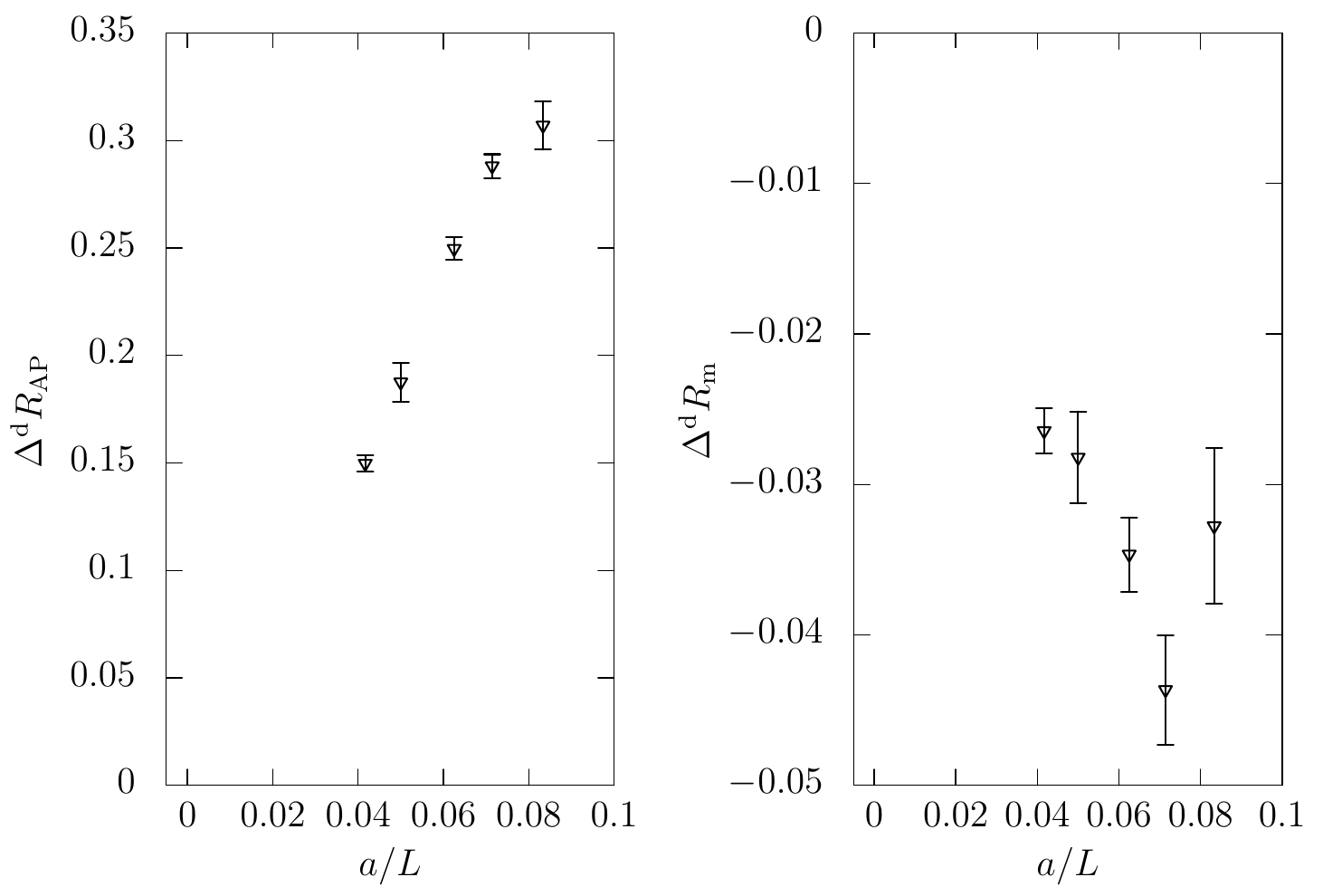}
    \caption{$\Or(a)$ ambiguities of $\RAP$ and $\Rm$ due to different
             definitions of the lattice derivative (improved vs. standard).
    }
    \label{img:amb_derivatives_rAP_m}
\end{figure}

\begin{figure}[t]
    \includegraphics[width=\linewidth]{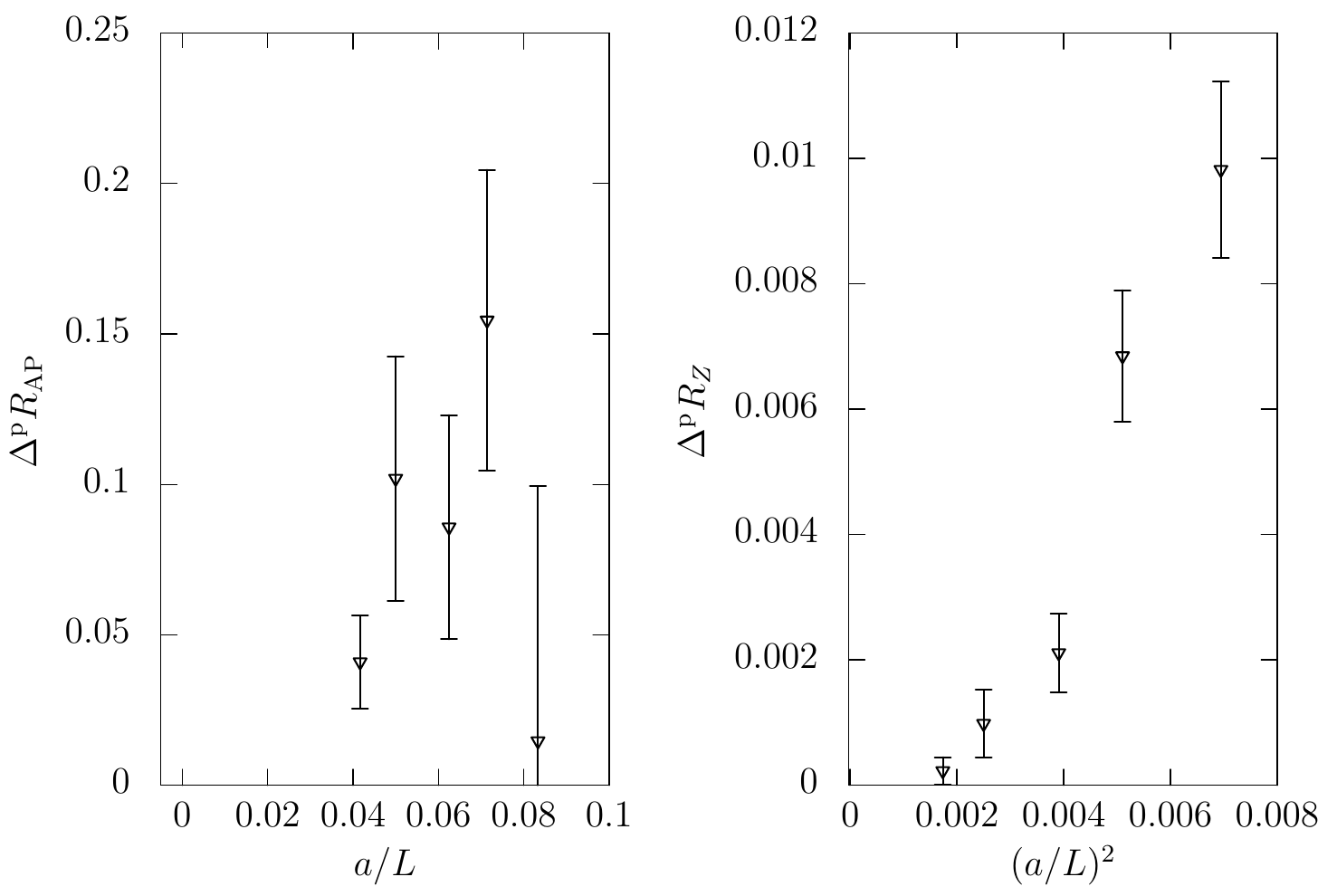}
    \caption{$\Or(a)$ ambiguity for $\RAP$ and $R_Z$, originating from third-
             vs.\ fourth-degree polynomial fits.
    }
    \label{img:amb_deg_rAP_Z}%
\end{figure}

\begin{figure}[t]
    \includegraphics[width=\linewidth]{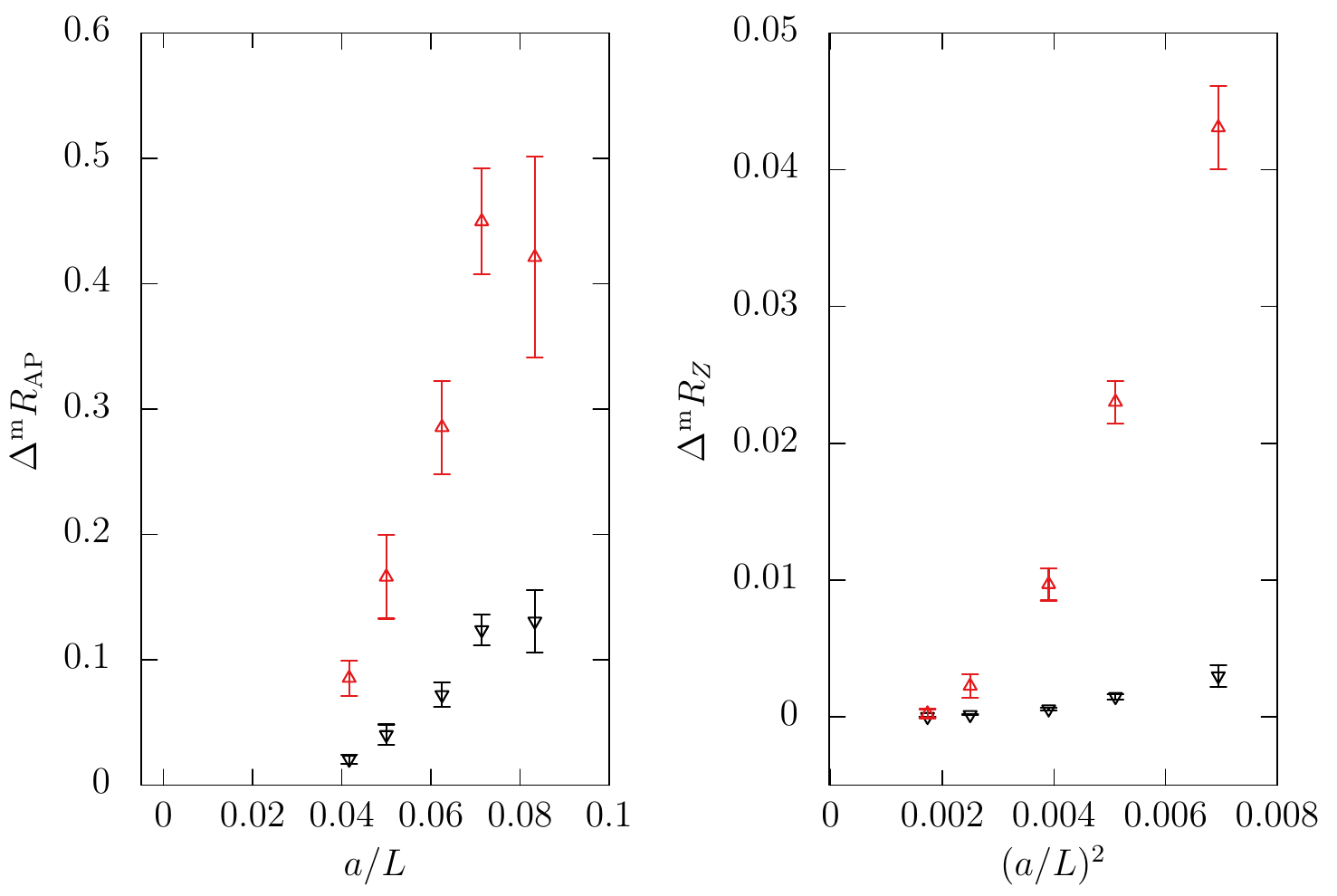}
    \caption{$\Or(a)$ ambiguities for $\RAP$ and $R_Z$ between different
             valence lines of constant physics. Black points show the
             differences between $L\Delta_{22}=0.25$ and $L\Delta_{22}=0$,
             while red points do so for $L\Delta_{22}=1$ and $L\Delta_{22}=0$.
    }
    \label{img:amb_masses_rAP_Z}%
\end{figure}

\subsubsection{Standard vs. improved derivatives}
\label{subsec:ambigderivatives}

In figure~\ref{img:amb_derivatives_rAP_m} we show the differences between final
results obtained using improved (``imp'') and standard (``std'') lattice
derivatives (cf.~eq.~\eqref{eqn:deriv})
\begin{align}
  \Delta^{\rm d} R_X \equiv
  \left. R_X \right|_\mathrm{imp} - \left. R_X \right|_\mathrm{std} \;,
\end{align}
for $\RAP$ and $\Rm$ in the LCP-0 case.
These arise as a consequence of $\Or(a)$ discretisation effects.
$\Delta^{\rm d} \Rm$ is of the order of the statistical errors. As found in the
quenched and two-flavour
analyses~\cite{deDivitiis:1997ka,Guagnelli:2000jw,Fritzsch:2010aw},
the estimator $\RAP$ is particularly sensitive to the chosen discretisation
of the derivatives, resulting in larger ambiguities. 
\mbox{Although} fluctuations are present, especially for the
largest lattice spacings, the $\Delta^{\rm d} R_X $ seem to vanish linearly in the
continuum limit as expected, see figure~\ref{img:amb_derivatives_rAP_m}. 

\subsubsection{Degree of the polynomial fits}
\label{subsec:ambigpolydegree}

Our results are obtained by fitting PCAC masses with third-degree
polynomials. The polynomial degree introduces a further source of
uncertainty, which we investigate by monitoring the quantity
\begin{align}
  \Delta^{\rm p} R_X \equiv R_X^{(\mathrm{deg}=3)} - R_X^{(\mathrm{deg}=4)} \;,
\end{align}
for our LCP-0 results, extrapolated to the chiral limit.
Figure~\ref{img:amb_deg_rAP_Z} illustrates the differences in $\RAP$ and $R_Z$.
Here, the resulting intrinsic ambiguities in $\RAP$ and $\Rm$ are $\Or(a)$,
while in $R_Z$ they are $\Or(a^2)$. These effects, which in case of 
$\RAP$ and $R_Z$ appear to be barely larger than their statistical errors,
vanish at smaller lattice spacings.
A qualitatively similar behaviour is observed for $\Rm$.

\subsubsection{Determination with non-unitary valence quark masses}
\label{subsec:ambigmass}

The polynomial fits allow the determination of the estimators at any valence
point in the considered mass range $0\leq L\Delta_{22} \leq 1$. Results
at different values of $L\Delta_{22}$ differ by mass-dependent cut-off effects.
Therefore, we also investigate the difference between the LCP-0 results
($L\Delta_{22}=0$) and two choices of heavy valence quarks, namely those at
a point with $L\Delta_{22}=0.25$ and at LCP-1 ($L\Delta_{22}=1$).
In figure~\ref{img:amb_masses_rAP_Z} we plot
\begin{align}
  \Delta^{\rm m} R_X \equiv
  \left. R_X \right|_{L\Delta_{22}>0} - \left. R_X \right|_{L\Delta_{22}=0}
\end{align}
for $\RAP$ and $R_Z$.
From the scaling behaviour of these estimators it is evident that the
relative size of the cut-off effects grows with the valence quark masses,
while the differences themselves decrease significantly towards the continuum 
limit.
(Note that this decrease is actually faster than the expected rates 
${\propto a/L}$ and $\propto a^2/L^2$ for improvement coefficients and 
renormalisation factors, respectively.)
As seen in figure~\ref{img:amb_masses_rAP_Z}, the difference 
$\Delta^{\rm m} R_X$ at fixed $a/L$ roughly scales with an integer power of the 
ratio of $L\Delta_{22}$, i.e., $1/4$ for $X={\rm AP}$ and $(1/4)^2$
for $X=Z$.  

We note in passing that there is also an implicit dependence of the results
on our choice of mass range $0\leq L\Delta_{22} \leq 1$.

\section{Results}
\label{sec:results}

Based on our non-perturbative calculation of the estimators $R_X^{(0)}$
listed in tables~\ref{tab:restab} and~\ref{tab:restab_Lx1}, we now provide
interpolating formulae to make them accessible also at other values of the
gauge coupling.

\begin{figure}[p]
        \centering
        \includegraphics[width=\textwidth]{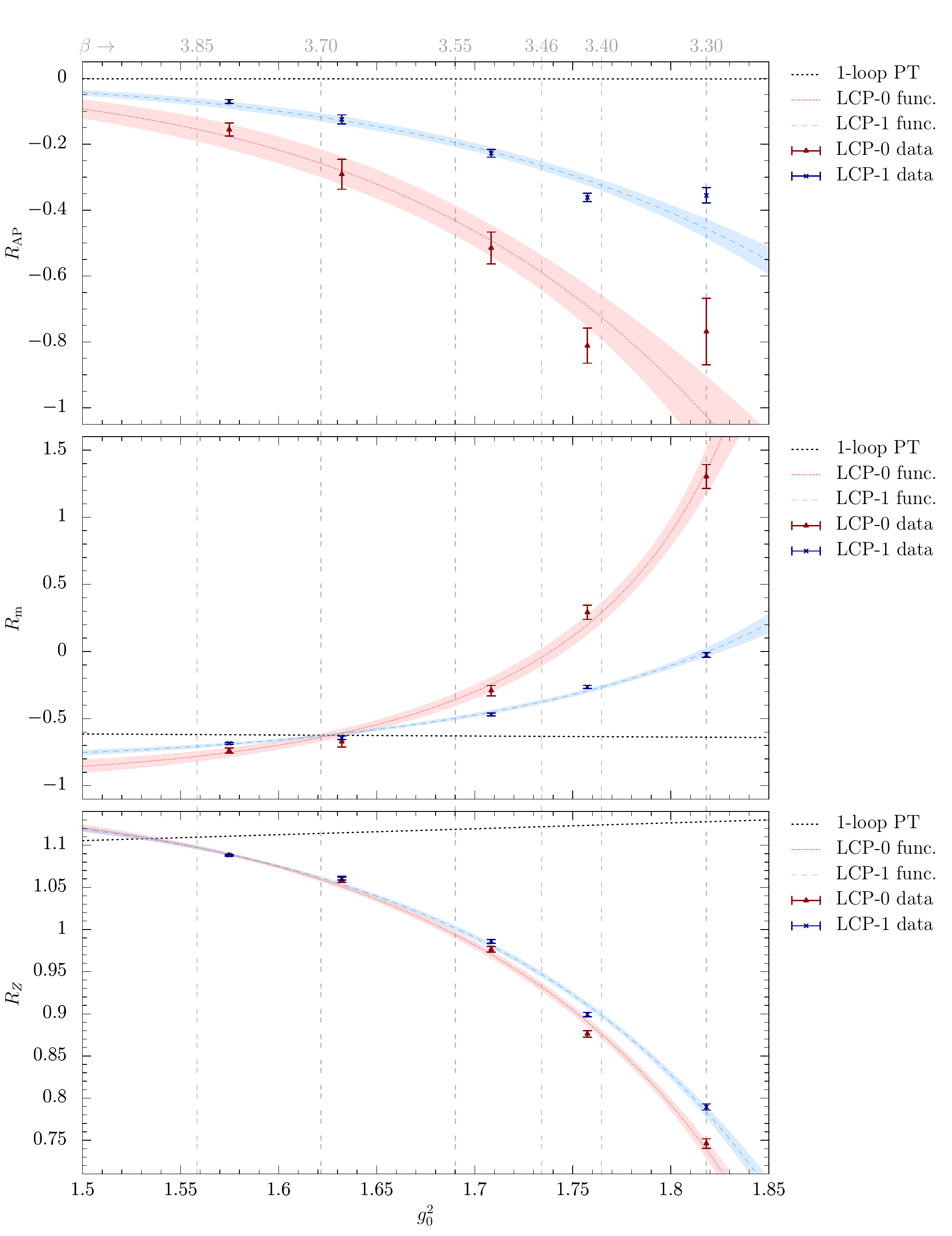}
        \caption{Estimators $R^{(0)}_X$ for LCP-0 and LCP-1 and
                 their interpolating functions according to
                 eqs.~\eqref{eq:RX_g0sq}.
                 Vertical lines indicate CLS $g_0^2$-values,
                 while straight dotted lines represent the
                 one-loop perturbative estimates.
                }
        \label{fig:g0sq-results}
\end{figure}

To have at least some constraint towards smaller couplings
$g_0^2=6/\beta$, we opt for interpolating formulae that encompass the 
one-loop perturbative behaviour as $g_0^2\to 0$.
Since there are no theoretical expectations for the functional forms of the
estimators in the region of large couplings, we have probed, with varied
degrees of success, many conceivable ans\"atze. We have settled for the
following ones:
\begin{subequations}\label{eq:RX_g0sq}
\begin{align}
\label{eq:RAP_g0sq}
  \RAP(g_0^2) &= \hphantom{-0.5}   -0.0010666 \,g_0^2 \times \left\{ 1+\exp\big( p_0 + p_1/g_0^2 \big) \right\}  \;, \\
\label{eq:Rm_g0sq}
   \Rm(g_0^2) &=             -0.5  -0.0762933 \,g_0^2 \times \frac{1+q_0\,g_0^2 +q_1\,g_0^4}{1+q_2\,g_0^2}        \;, \\
\label{eq:RZ_g0sq}
   R_Z(g_0^2) &= \hphantom{-} 1.0  +0.0703169 \,g_0^2 \times \frac{1+z_0\,g_0^2 +z_1\,g_0^4}{1+z_2\,g_0^2}        \;.
\end{align}
\end{subequations}
The numerical constants in the above equations are those dictated by one-loop
perturbation theory~\cite{Aoki:1998ar,Taniguchi:1998pf}.
The other parameters are determined from fits. 
At the unitary chiral point (corresponding to LCP-0) we obtain:
\begin{subequations}\label{eq:RX_LCP0_par}
\begin{align}
    (p_j)  &= \left( 16.7457,  -19.0475               \right)  \;, \\
    (q_j)  &= \left( 3.53337,  -2.48944,  -0.516695  \right)  \;,  \\
    (z_j)  &= \left( 0.703413, -0.769835, -0.478372  \right)  \;,        
\end{align}
\end{subequations}
with covariance matrices
\begin{subequations}\label{eq:RX_LCP0_cov}
\begin{align}
	\mathrm{cov}(p_i,p_j) &= \begin{pmatrix*}[r]
                                 3.49591 & -6.07560 \\
                                -6.07560 & 10.5834 \\         
                        	 \end{pmatrix*}    \;,\\
        \mathrm{cov}(q_i,q_j) &= \begin{pmatrix*}[r]
                                 94.5681 & -57.5056 &  0.859064 \\
                                -57.5056 &  34.9883 & -0.525367 \\
                                  0.859064 &  -0.525367 &  0.009086 \\
                        	 \end{pmatrix*} \times 10^{-2}    \;,\\
        \mathrm{cov}(z_i,z_j) &= \begin{pmatrix*}[r]
                                 4.22703 & -2.54941 &  0.231607   \\
                                -2.54941 &  1.537772 & -0.139695   \\
                                 0.231607 & -0.139695 &  0.013179   \\
                                 \end{pmatrix*} \times 10^{-2}    \;.
\end{align}
\end{subequations}
For the estimators in the partially-quenched setup, at a fixed physical
heavy valence quark mass (corresponding to LCP-1), we find:
\begin{subequations}\label{eq:RX_LCP1_par}
\begin{align}
    (p_j)  &= \left( 15.6049, -18.4592              \right)  \;,  \\
    (q_j)  &= \left( 2.66968, -1.93055,  -0.468542  \right)  \;,  \\
    (z_j)  &= \left( 0.729908, -0.780933, -0.467403  \right)  \;,  
\end{align}
\end{subequations}
with covariance matrices
\begin{subequations}\label{eq:RX_LCP1_cov}
\begin{align}
	\mathrm{cov}(p_i,p_j) &= \begin{pmatrix*}[r]
                                 1.50497 & -2.63930   \\
                                -2.63930 &  4.63683   \\
                        	 \end{pmatrix*}     \;, \\
  	\mathrm{cov}(q_i,q_j) &= \begin{pmatrix*}[r]
                                 74.2042 & -44.4131 &  2.10959 \\
                                -44.4131 &  26.5860 & -1.26398 \\
                                  2.10959 & -1.26398 &  0.062898 \\
                        	 \end{pmatrix*} \times 10^{-2}    \;,\\
        \mathrm{cov}(z_i,z_j) &= \begin{pmatrix*}[r]
                                 2.94708 & -1.76762 &  0.182059   \\
                                -1.76762 &  1.06029 & -0.109193   \\
                                 0.182059 & -0.109193 &  0.011597   \\         
                                 \end{pmatrix*} \times 10^{-2}    \;.
\end{align}
\end{subequations}
These continuous parameterisations of the results, together with the data,
are presented in figure~\ref{fig:g0sq-results}.
For reasons explained below, the covariance matrices are inflated
by about a factor of two.

We have also produced a gauge configuration ensemble at $\beta=8.0$, 
in order to evaluate the estimators $R_X$ in the deeply perturbative region.
Even though the physical volume of this ensemble is smaller than the LCP
one, these $\beta=8.0$ results qualitatively agree with our fit ans\"atze
and their shape as $g_0^2\to 0$. This corroborates our
interpolating fit functions, which asymptotically approach the one-loop
perturbative predictions at small gauge couplings.

\begin{table}[t]%
	\centering\small
	\renewcommand{\arraystretch}{1.25}
	\begin{tabular}{CLLLLLL}
\toprule
& \multicolumn{3}{c}{LCP-0} & \multicolumn{3}{c}{LCP-1} \\\cmidrule(lr){2-4}\cmidrule(lr){5-7}
 \beta &  \RAP      &  \Rm       &  R_Z       &  \RAP      &  \Rm       &  R_Z         \\\midrule
 3.85  & -0.155(36) & -0.781(38) & 1.0975(25) & -0.073(12) & -0.708(15) & 1.0971(18)   \\
 3.70  & -0.258(42) & -0.640(31) & 1.0591(23) & -0.119(14) & -0.630(11) & 1.0612(17)   \\
 3.55  & -0.432(46) & -0.358(47) & 0.9937(42) & -0.196(14) & -0.498(15) & 1.0015(30)   \\
 3.46  & -0.590(53) & -0.044(65) & 0.9320(50) & -0.265(14) & -0.376(17) & 0.9468(35)   \\
 3.40  & -0.726(67) & +0.290(76) & 0.8758(52) & -0.324(17) & -0.266(17) & 0.8981(35)   \\
\bottomrule
\end{tabular}

%3.40  & -0.732(37)		&	\phantom{-}0.383(35)	&	0.8687(25) \\
%3.46  & -0.638(31)		&	\phantom{-}0.003(29)	&	0.9285(23) \\
%3.55  & -0.504(23)		&	-0.418(31)	&	0.9980(24) \\
%3.70  & -0.294(17)		&	-0.743(24)	&	1.0677(18) \\
%3.85  & -0.101(22)(53)	&	-0.687(26)(95)	&	1.0901(16)(74) \\

%#---------------------------------------------------------
%#beta  obs  LCP-0                    LCP-1                
%#---------------------------------------------------------
% 3.85  RZ   +1.097459 +/- 0.002470  +1.097127 +/- 0.001823
% 3.70  RZ   +1.059109 +/- 0.002291  +1.061176 +/- 0.001679
% 3.55  RZ   +0.993724 +/- 0.004237  +1.001530 +/- 0.003006
% 3.46  RZ   +0.932012 +/- 0.005028  +0.946806 +/- 0.003453
% 3.40  RZ   +0.875809 +/- 0.005221  +0.898137 +/- 0.003496
% 3.30  RZ   +0.738941 +/- 0.010610  +0.783784 +/- 0.006844
% 3.85  Rm   -0.781163 +/- 0.038292  -0.707890 +/- 0.015034
% 3.70  Rm   -0.640106 +/- 0.031356  -0.630071 +/- 0.011031
% 3.55  Rm   -0.358272 +/- 0.047181  -0.498376 +/- 0.014726
% 3.46  Rm   -0.043597 +/- 0.065384  -0.375899 +/- 0.016690
% 3.40  Rm   +0.289697 +/- 0.075963  -0.266058 +/- 0.017073
% 3.30  Rm   +1.345135 +/- 0.170348  -0.005467 +/- 0.036008
% 3.85  RAP  -0.154839 +/- 0.036108  -0.072936 +/- 0.011734
% 3.70  RAP  -0.258321 +/- 0.042034  -0.119422 +/- 0.013500
% 3.55  RAP  -0.432334 +/- 0.045506  -0.196466 +/- 0.013797
% 3.46  RAP  -0.589651 +/- 0.052723  -0.265344 +/- 0.014271
% 3.40  RAP  -0.725558 +/- 0.067011  -0.324430 +/- 0.016768
% 3.30  RAP  -1.026102 +/- 0.121622  -0.454072 +/- 0.029938
%#---------------------------------------------------------

    \caption{Interpolated values of our estimators for couplings employed 
             in CLS simulations along the two renormalised trajectories 
             LCP-0 and LCP-1 considered in this work.
             Statistical uncertainties are as described in the text and
             match the confidence band in figure~\ref{fig:g0sq-results}.
            }
	\label{tab:clstab}
\end{table}
We cover the $g_0^2$-range typical of large-volume computations of bare quark 
masses, matrix elements and other phenomenological applications. 
In particular, the $\Nf = 2+1$ couplings of the CLS effort in 
refs.~\cite{Bruno:2014jqa,Bruno:2016plf,Bali:2016umi,Bruno:2017gxd,Mohler:2017wnb} 
are ${\beta\in\{3.85,3.70,3.55,3.46,3.4\}}$.
In figure~\ref{fig:g0sq-results} we indicate the CLS $g_0^2$-values as
vertical dashed lines, and in table~\ref{tab:clstab} we provide our
interpolated results at the corresponding $\beta$-values. 
Note that the smallest coupling employed in the CLS simulations,
being slightly outside our range ${\beta \in [3.3,3.81]}$, can be reached by
extrapolation of our interpolating functions. In these cases, near the edges
of our $\beta$-range, the results are more sensitive to the choice of the
specific fit ansatz. 
The covariance matrices of eqs.~(\ref{eq:RX_LCP0_cov}) 
and (\ref{eq:RX_LCP1_cov}), as well as the statistical errors in
table~\ref{tab:clstab}, are large enough to cover the results obtained from 
different fit ans\"atze we have tried out.
Moreover, it can be seen from figure~\ref{fig:g0sq-results} that the error 
band at the CLS couplings is consistent with the errors of the neighbouring 
data points.
We have verified explicitly that any effect on the error of a typical
combination of our estimators coming from the correlations among the $R_X$'s
(which in principle would have to be taken into account when, for instance,
calculating renormalised quark masses) is negligible compared to the 
inflated statistical uncertainties of the fits.
Nevertheless, for the sake of completeness, we quote in 
appendix~\ref{app:correlations} the correlation matrices among the $R_X$'s.

\begin{figure}[t]
	\centering
	\includegraphics[width=.75\linewidth]{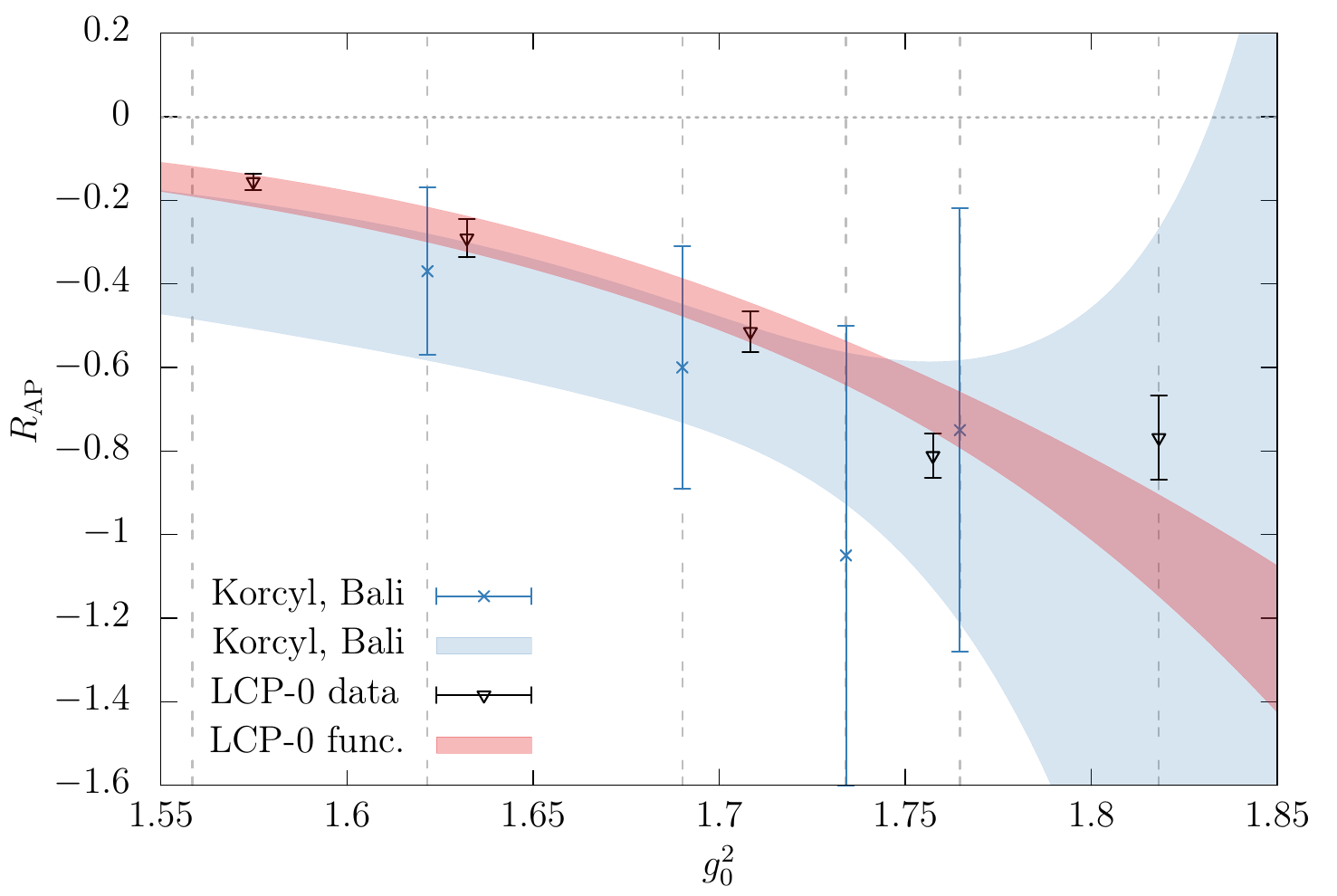}
	\caption{The chirally extrapolated estimator $\smash{\RAP^{(0)}}$
                 (in the sector of trivial topology) as a function of the
                 squared coupling, $g_0^2$, together with the values and the
                 curve determined by Korcyl and Bali in~\cite{Korcyl:2016ugy}.
		 The dashed line shows the one-loop perturbative prediction.
                 The vertical lines indicate the couplings used in the CLS
                 large volume simulations
                 \cite{Bruno:2014jqa,Bruno:2016plf,Bali:2016umi}.}
	\label{img:R_AP_kb}
\end{figure}

Our results are compatible with the recent ones by Korcyl and
Bali~\cite{Korcyl:2016ugy} within their considerably larger errors;
cf.~figure~\ref{img:R_AP_kb}.
Any differences between the two sets of results at the same coupling
$g_0^2$ are to be attributed to discretisation effects.

Finally we note that, since 
$Z_\mathrm{S}/Z_\mathrm{P}=(Z_\mathrm{m}Z_\mathrm{P})^{-1}=(Z_\mathrm{A}Z)^{-1}$,
the (scale independent) ratio of renormalisation constants 
$Z_\mathrm{S}/Z_\mathrm{P}$ may be obtained by combining $R_Z$ from this work 
with the interpolation formula for $Z_\mathrm{A}$ 
from~\cite{Bulava:2016ktf,DallaBrida:2018tpn}.
A direct determination of $Z_\mathrm{S}/Z_\mathrm{P}$ based on the Ward
identity approach is in progress~\cite{Heitger:2018pwb}.

\section{Conclusions}
\label{sec:concl}

The present paper is part of a series of publications dedicated to the
non-perturbatively $\rmO(a)$ improved quark mass renormalisation in
three-flavour lattice QCD with Wilson fermions.
It complements previous determinations of the axial current improvement and
normalisation~\cite{Bulava:2015bxa,Bulava:2016ktf,DallaBrida:2018tpn} and
the renormalisation factor of the pseudoscalar density~\cite{Campos:2018ahf}
by a non-perturbative calculation of the improvement coefficients $\ba-\bp$
and $\bm$ --- multiplying associated additive, quark mass dependent Symanzik
counter-terms --- as well as of the normalisation factor
$Z\equiv Z_m\zp/\za$.
We work in the framework of lattice QCD with $\Nf=3$ flavours
of mass-degenerate, non-perturbatively $\rmO(a)$ improved Wilson-clover
sea quarks and tree-level Symanzik-improved gluons.

Our computational setup to determine $\ba-\bp$, $\bm$ and $Z_m\zp/\za$
consists in small physical volume simulations, with Schr\"odinger functional
boundary conditions, and exploiting the PCAC relation with mass
non-degenerate valence quarks.
Valence quark masses and lattice volumes have been varied ensuring that
we approach the chiral and continuum limits while staying on a line of
constant physics.
Although we have based our work on an earlier $\Nf=2$
publication~\cite{Fritzsch:2010aw}, we have extended that analysis by
introducing a series of novelties as explained in the main part of the
paper.
The final results obtained refer to massless sea quarks.
These can be inferred from tables~\ref{tab:restab} and \ref{tab:restab_Lx1},
together with the 
formulae~(\ref{eq:RX_g0sq}), (\ref{eq:RX_LCP0_par}) and 
(\ref{eq:RX_LCP1_par}), which provide smooth parameterisations of
$\ba-\bp$, $\bm$ and $Z_m\zp/\za$ in terms of the bare gauge coupling
squared.
Several checks have been performed to address the various systematics
involved and to guarantee the stability of the analysis strategy as well as 
the reliability of the quoted error estimates.

Since our range of couplings matches that of the large-volume CLS 
simulations~\cite{Bruno:2014jqa,Bruno:2016plf,Bali:2016umi,Mohler:2017wnb}, 
our results are currently being applied in a $(2+1)$-flavour computation of 
light, strange and charm quark masses 
(see~\cite{Bruno:2019xed,Heitger:2019ioq} for a preliminary account).
They are also useful in the computation of other physical quantities, 
such as certain combinations of QCD matrix elements involving the axial 
current and the pseudoscalar density. 
Our results for $\ba-\bp$, $\bm$, and $Z$ for the specific bare couplings 
of the CLS ensembles are collected in table~\ref{tab:clstab}.

\begin{acknowledgement}
We wish to thank Maurizio Firrotta for participating in the early stages of 
this project. A.\,V. wishes to thank Steve Sharpe for illuminating discussions 
and the University of M\"unster for its hospitality. 
We also like to thank Christian Wittemeier, Fabian Joswig and Carlos Pena for
many useful discussions and particularly Fabian for his valuable contributions
in extending the set of ensembles used in our computations.

Computer resources were provided by the ZIV of the University of Münster 
(PALMA, PALMA-NG and PALMA-II HPC clusters), the INFN (GALILEO cluster at 
CINECA) and the HTCondor Cluster of the Institut für Theoretische Physik 
of the University of Münster.

This work is supported by the Deutsche Forschungsgemeinschaft (DFG) through 
the Research Training Group \textit{“GRK 2149: Strong and Weak Interactions 
- from Hadrons to Dark Matter”} (J.\,H. and S.\,K.).
C.\,C.\,K., scholar of the German Academic Scholarship Foundation 
(Studienstiftung des deutschen Volkes), gratefully acknowledges their 
financial and academic support.

\end{acknowledgement}

\clearpage
\appendix
\section{Non-unitary QCD and improvement coefficients}
\label{app:pQCD}

In this appendix we will further discuss how the key expressions of sect.~\ref{sec:renormimpr} are modified in the non-unitary version of lattice QCD with sea quarks with masses $m_{{\rm q},i} (i = 1, \ldots , \NF)$ and valence quarks with masses $m_{{\rm q},i}^{\rm val} (i = 1, \ldots , \NVAL)$. It is understood that both sea and valence lattice fermion actions are regularised {\it \`a la} Wilson. In general $\NVAL \neq \NF$ and sea and valence quark masses are unequal. The chiral limit $\kappa_{\rm crit}$ is the one defined, in some standard fashion, in the unitary theory of $\NF$ 
quarks. 
All subtracted sea quark masses are defined so as to vanish at $\kappa_{\rm crit}$.

The renormalisation and improvement pattern of each valence quark mass 
$[m_i^{\rm val}]_{\rm R}$ depends on the sea quark mass matrix $\Tr{\Mq}$ and the bare 
quark mass $m^{\rm val}_{{\rm q},i}$ of this very flavour.
(With some standard renormalisation condition imposed for the quark mass, there is no 
physical reason for a dependence on a valence quark mass of a different valence 
flavour.) This leads to the expression
\begin{align}
[m_i^{\rm val}]_{\rm R}
&= \zm^\prime \bigg \{ \Big [\, m_{{\rm q},i}^{\rm val} + k_{\rm m} \dfrac{\Tr{\Mq}}{\NF} \,\Big ] 
\label{eq:M-imp-pqf} \\
&\quad + a\,\Big [\, h_{\rm m} (m_{{\rm q},i}^{\rm val})^2  + \bar h_{\rm m} m_{{\rm q},i}^{\rm val} \Tr{\Mq} + j_{\rm m} \dfrac{\Tr{\Mq^2}}{\NF} + \bar j_{\rm m} \dfrac{\Tr{\Mq}^2}{\NF} \,\Big ] \bigg \} \;.
\nonumber
\end{align}

To the order we are working in the lattice spacing, the coefficients $k_{\rm m}$, $h_{\rm m}$, $\bar h_{\rm m}$, $j_{\rm m}$ and $\bar j_{\rm m}$ depend on the bare coupling $g_0^2$  only; any mass dependence is an $\rmO(a\mq)$  discretisation effect. If we drive the valence bare quark mass to the value of the corresponding sea quark mass  (i.e., $\kappa_i^{\rm val} = \kappa_i$), the above expression should reduce to eq.~(\ref{eq:mren-mq}). 
This implies the identification
\begin{align}
\zm^\prime &= \zm \;, \qquad && k_{\rm m} = r_{\rm m} - 1 \;, \\
h_{\rm m}  &= \bm \;, \qquad && \bar h_{\rm m} = \bar b_{\rm m} \;, \\
j_{\rm m}  &= r_{\rm m} d_{\rm m} - \bm \;, \qquad && \bar j_{\rm m} = r_{\rm m} \bar d_{\rm m} - \bar b_{\rm m}  \;,
\end{align}
and eq.~(\ref{eq:M-imp-pqf}) becomes
\begin{align}
[m_i^{\rm val}]_{\rm R}
&= \zm \bigg \{ \Big [\, m_{{\rm q},i}^{\rm val} 
+ (r_{\rm m} -1) \dfrac{\Tr{\Mq}}{\NF} \,\Big ] 
\label{eq:M-imp-pqf1} \\
&\quad + a\,\Big [\, \bm (m_{{\rm q},i}^{\rm val})^2  + \bar b_{\rm m} m_{{\rm q},i}^{\rm val} \Tr{\Mq}
\nonumber \\
&\quad + (r_{\rm m} d_{\rm m} - \bm) \dfrac{\Tr{\Mq^2}}{\NF} + (r_{\rm m} \bar d_{\rm m} - \bar b_{\rm m})  \dfrac{\Tr{\Mq}^2}{\NF} \,\Big ] \bigg \} \;.
\nonumber
\end{align}
This is simply  eq.~(\ref{eq:mren-mq}), with $\mqi$ denoting valence quark masses and $\Mq$ the mass matrix of sea quark masses. 
Following the same reasoning, we conclude that analogous results hold for 
eqs.~(\ref{eqn:ren-off-diag-A}), (\ref{eqn:ren-off-diag-P}), (\ref{eq:renmassPCAC}) and (\ref{eq:massmatch}).

As a quick cross-check, we trace the last expression over all valence flavours:
\begin{align}
\Tr{M_{\rm val}}_{\rm R}
&= Z_m \bigg \{ \Big [\, \Tr{M_{\rm val}}  + (r_m -1) \dfrac{\Tr{\Mq}}{\NF} \NVAL \,\Big ] 
\label{eq:M-imp-pqf2} \\
&\quad + a\,\Big [\, b_m \Tr{M_{\rm val}^2}  + \bar b_m \Tr{M_{\rm val}} \Tr{\Mq}
\nonumber \\
&\quad + (r_m d_m - b_m) \dfrac{\Tr{\Mq^2}}{\NF} \NVAL + (r_m \bar d_m - \bar b_m)  \dfrac{\Tr{\Mq}^2}{\NF} \NVAL \,\Big ] \bigg \} \;.
\nonumber
\end{align}
We then drive the ``non-unitary QCD'' formulation to the unitary QCD one: 
This means that $\NVAL = \NF$, and for each flavour the valence quark mass
is equal to that of the sea. The above expression reduces to eq.~(25) of ref.~\cite{Bhattacharya:2005rb}.
Similarly we obtain
\begin{equation}
\Tr{\lambda^a_{\rm val} M_{\rm val}}_{\rm R} \, = \, \zm \Big [\, (1 + a \bar b_m \Tr{\Mq}) \Tr{\lambda^a_{\rm val} M_{\rm val}} + a b_m  \Tr{\lambda^a_{\rm val} M_{\rm val}^2} \,\Big ] \;,
\label{eq:M-imp-ns-pqcd}
\end{equation}
which is eq.~(24) of ref.~\cite{Bhattacharya:2005rb} when the non-unitary formulation is driven to the unitary one.

\section{Schr\"odinger functional correlation functions}
\label{app:SF}

Following ref.~\cite{Luscher:1996sc}, we define the operator $\mathcal{O}^{ji}$ of eq.~(\ref{eqn:PCAC-mass}) in terms of the boundary quark and anti-quark fields $\zeta^i$ and $\zetabar^j$ at Euclidean time $x_0=0$ (and also operator the $\mathcal{O}^{\prime ji}$ in terms of the boundary fields $\zeta^{\prime j}$ and $\zetabar^{\prime i}$ at Euclidean time $x_0=T$):
\begin{align}        
\mathcal{O}^{ji}
&= 
\dfrac{a^6}{L^3}\sum_{\vecu,\vecv}  
\zetabar^j(\vecu)\,\dirac5\,\zeta^i(\vecv) \;, &
\mathcal{O}^{\prime ji} 
&= 
\dfrac{a^6}{L^3}\sum_{\vecu,\vecv}  
\zetabarprime^j(\vecu)\,\dirac5\,\zetaprime^i(\vecv) \;.
\label{eqn:PS-ops}
\end{align}
Summed over the spatial volume, these yield pseudoscalar boundary sources projected onto zero momentum.
From these, the $x_0 = 0$ boundary-to-bulk forward Schr\"odinger functional 
(SF) correlation functions in the pseudoscalar channel are constructed from 
the axial current and density as 
\begin{align}
\fa^{ij}(x_0) 
&=
-\dfrac{a^3}{2}\sum_{\vecx}\big\langle A_0^{ij}(x)\,\ob^{ji}\,\big\rangle \;, &
\fp^{ij}(x_0) 
&=
-\dfrac{a^3}{2}\sum_{\vecx}\,\big\langle P^{ij}(x)\,\ob^{ji}\,\big\rangle \;.
\label{eqn:def-fA-fP-ij}
\end{align}
Flavour indices $i,j$ are not summed over, and when $i=j$ they denote degenerate but distinct flavours. With $\mathcal{O}^{ji}$ replaced by $\mathcal{O}^{\prime ji}$, we also have the  $x_0=T$ boundary-to-bulk backward SF correlation functions $g^{ij}_{\rm A, P}(T-x_0)$. In a vanishing background field, they are related to $f^{ij}_{\rm A, P}(x_0) $ by time reflection and are averaged in order to reduce the statistical noise. 

In our SF framework, the bare PCAC quark masses of eq.~(\ref{eqn:PCAC-mass}) are given by:
\begin{align}
m_{ij}(x_0)\equiv m_{ij}(x_0;L/a,T/L,\theta)
&= 
\dfrac{\sdrv0\fa^{ij}(x_0)+a\ca\drvstar{0}\drv{0}\fp^{ij}(x_0)}
{2\,\fp^{ij}(x_0)} \;,
\label{eqn:mpcac_x0-ij}
\end{align}
where we explicitly indicate their additional dependence on $L/a$, $T/L$ and the periodicity angle $\theta$ in the boundary conditions of the fermion fields. These dependences will usually be implicit, in order to keep the notation simple. In the degenerate case ($i=j$), $m_{ij}$ reduces to the non-singlet PCAC mass of a flavour degenerate doublet. 

The first and second lattice derivatives $\sdrv0$ and $\drvstar{0} \drv{0}$ in the last equation, upon acting on smooth functions, are the continuum ones up to terms of $\Or\big(a^2\big)$ and $\Or\big(a\big)$, respectively. Following refs.~\cite{deDivitiis:1997ka,Guagnelli:2000jw}, besides using these derivatives, we have computed current quark masses involving derivatives obtained with the replacements
\begin{align}
\sdrv0
&\rightarrow
\sdrv0\left(1-{\T \frac{1}{6}}\,a^2\drvstar{0}\drv{0}\right) \;, &
\drvstar{0}\drv{0}
&\rightarrow
\drvstar{0}\drv{0}
\left(1-{\T \frac{1}{12}}\,a^2\drvstar{0}\drv{0}\right) \;.
\label{eqn:deriv}
\end{align}
Upon acting on smooth functions, these derivatives are the continuum ones up to terms of $\Or\big(a^4\big)$; thus they are ``improved'' as far as their discretisation effects are concerned. It is hoped that, when used in the definition of $m_{ij}$, the resulting estimates of $\bm, \ba - \bp$, and $Z$ will show milder discretisation effects. This is of course not guaranteed, as other terms of $\Or(a^2)$ from the correlation functions remain uncancelled.

% NOTE: Properly adjust explicit reference labels in the title to those of
% \cite{Guagnelli:2000jw,Fritzsch:2010aw} from the actual list of references
%
\section{Results from the method of refs.~[11,13]} 
\label{app:oldstrategy}

In refs.~\cite{Guagnelli:2000jw,Fritzsch:2010aw}, estimates for $R_X$ were 
obtained in the quenched and two-flavour cases respectively,
using a different method to analyse the data. Here we compare results from 
this method to those obtained from our analysis.

In ref.~\cite{Fritzsch:2010aw} the heavier mass was tuned to a single value,
$Lm_{22} \approx 0.50$. This was small enough to ensure small discretisation
effects and large enough to keep statistical uncertainties under control. Thus,
the results quoted in ref.~\cite{Fritzsch:2010aw} for $b_{\rm A}-b_{\rm P}$,
$\bm$ and $Z$ contain $\mathcal O(a m_{22})$ effects as part of their
non-perturbative definition. In the present work, with several $L m_{22}$
values at our disposal, we can extrapolate first to the unitary point
$m_{22} \rightarrow m_{11}$ and then to the chiral limit $m_{11} \rightarrow 0$.

\begin{figure}[t]
	\includegraphics[width=\linewidth]{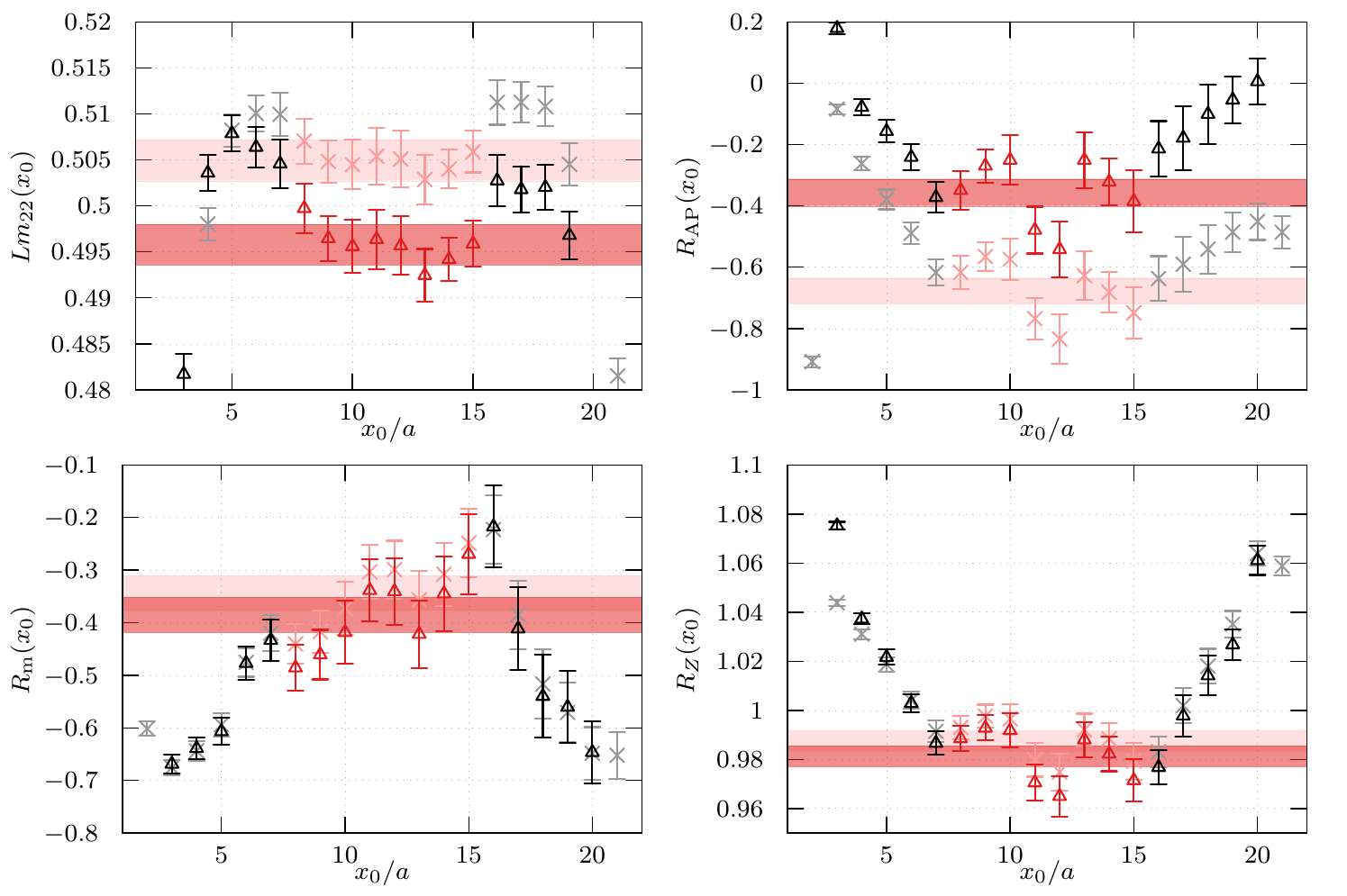}
    \caption{Results for the ensemble B1k4 and valence quarks with
             hopping parameter ${\kappa_2=0.13594}$, corresponding to
             $L\Delta_{22}\approx 0.5$. \textit{Top left panel}: PCAC mass
             $Lm_{22}$; the red points of the central third of the time
             extension are averaged to give the plateau, drawn as a red band.
             The points in the background with lighter colours (crosses) 
             correspond to the result from standard derivatives, while the 
             data points in the foreground with darker colours (triangles) 
             are derived from improved lattice derivatives.
             \textit{Other panels}: The $x_0$-dependence of the
             estimators $\RAP, \Rm$ and $R_Z$, and the associated
             plateau regions.
	        }
	\label{img:B1k4_plateau}
\end{figure}
\begin{figure}[h]
	\centering
	\includegraphics[width=.75\linewidth]{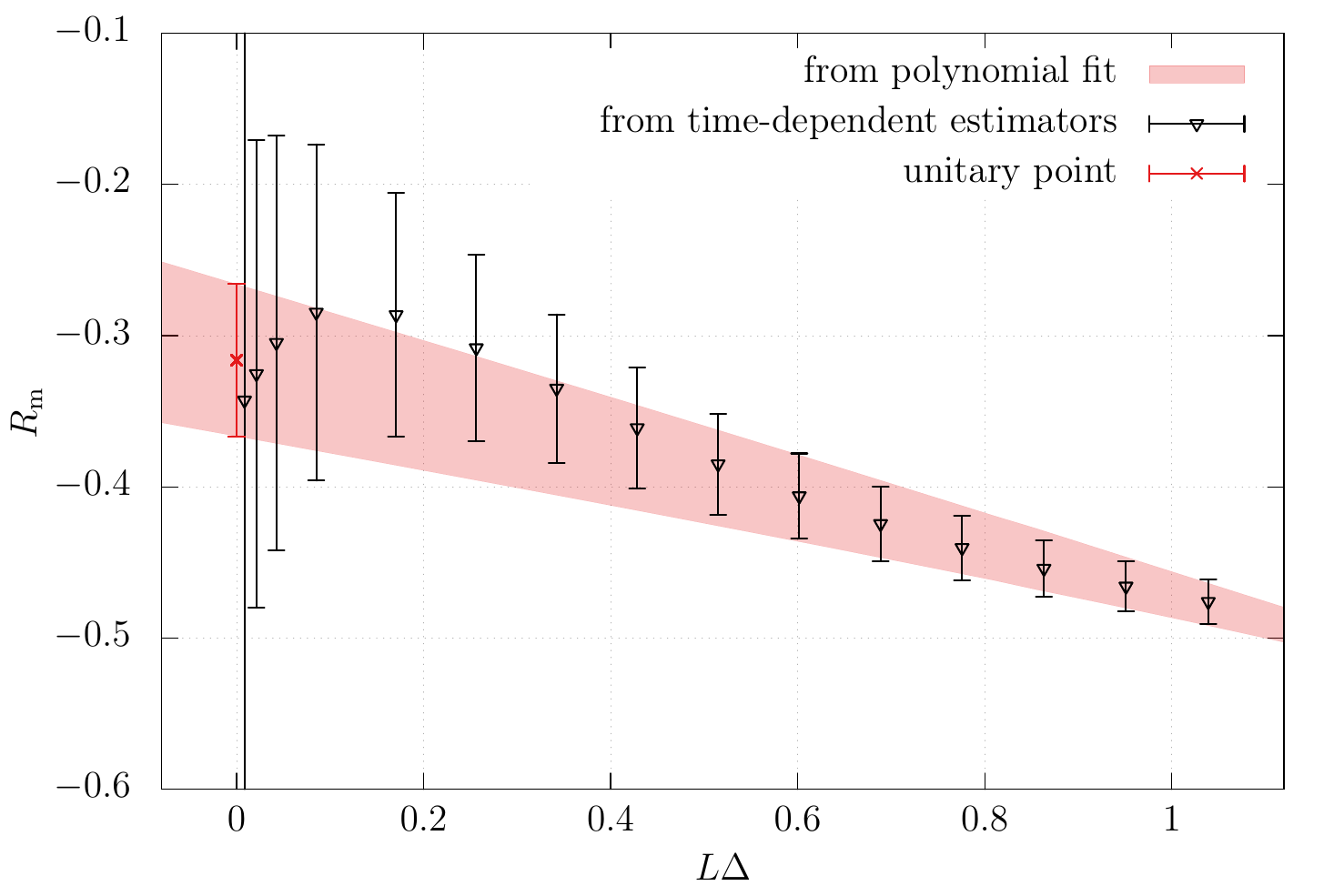}
	\caption{
		$\Rm$ plotted against $L\Delta$ for ensemble B1k4. 
                Points are obtained by measuring current quark masses directly 
                and using (\ref{eqn:estim-RZ}), as detailed in this appendix. 
                The continuous band results from the combined fits described 
                in subsection~\ref{subsec:estimators}. 
                The red point gives the result at the unitary point.
	}
	\label{img:Rfrommassfit}
\end{figure}
In the spirit of refs.~\cite{Guagnelli:2000jw,Fritzsch:2010aw} the current
masses $m_{11}, m_{12}, m_{22}$ and $m_{33}$ are computed at each time-slice
$x_0$ and fed into the definitions of the estimators $\RAP$, $\Rm$ and $R_Z$;
cf.~eqs.~(\ref{eqn:estim-RAP})--(\ref{eqn:estim-RZ}).
These, in theory, should also display plateaux as functions of $x_0$, being
functions of the current quark masses. However, as seen in
figure~\ref{img:B1k4_plateau}, this is not usually the case. 
As anticipated in subsect.~\ref{subsec:newstrategy}, the problem arises from 
numerical instabilities owing to the subtlety in the cancellation of nearly 
equal masses (such as $2 m_{12}$ and $m_{11} + m_{22}$ in $\RAP$).

We conclude that a change of strategy is required, in order to obtain
stable results when approaching unitarity.
In figure~\ref{img:Rfrommassfit}, the continuous band for $\Rm$, based on the
polynomial fits to the PCAC masses, is shown for comparison.
Results from both determinations agree for large valence quark masses.
Close to the unitary point, where the older method fails, the polynomial
fits give a safe $\Rm$-estimate.
A similar behaviour is seen for the other estimators.

\section{Results at $m_{11}\neq 0$}
\label{app:results}

Our results for current quark masses, $\RAP$, $\Rm$ and $R_Z$ are listed 
in tables~(\ref{tab:resAllEns_Q0})--(\ref{tab:resAllEns_Qall_Lx1}) for 
each configuration ensemble.
\begin{table}%
	\renewcommand{\arraystretch}{1.25}
	\centering\small
	\begin{tabular}{cllll}
\toprule
ID & $m_{11}^{(0)}$ & $R_\text{AP}^{(0)}$ & $R_\text{m}^{(0)}$ & $R_Z^{(0)}$\\
\midrule
A1k1  &  $-0.00278$$(80)$  &  $-1.016$$(119)$  &  $\phantom{+}1.344$$(157)$  &  $\phantom{+}0.7463$$(83)$\\
A1k3  &  $\phantom{+}0.00079$$(118)$  &  $-0.866$$(156)$  &  $\phantom{+}1.192$$(141)$  &  $\phantom{+}0.7420$$(96)$\\
A1k4  &  $-0.00110$$(36)$  &  $-0.744$$(89)$  &  $\phantom{+}1.395$$(73)$  &  $\phantom{+}0.7490$$(48)$\\
\hline
E1k1  &  $\phantom{+}0.00262$$(26)$  &  $-0.755$$(84)$  &  $\phantom{+}0.313$$(66)$  &  $\phantom{+}0.8687$$(63)$\\
E1k2  &  $-0.00022$$(22)$  &  $-0.816$$(57)$  &  $\phantom{+}0.290$$(57)$  &  $\phantom{+}0.8768$$(42)$\\
\hline
B1k1  &  $\phantom{+}0.00549$$(21)$  &  $-0.352$$(61)$  &  $-0.247$$(45)$  &  $\phantom{+}0.9676$$(48)$\\
B1k2  &  $\phantom{+}0.00444$$(25)$  &  $-0.271$$(90)$  &  $-0.272$$(73)$  &  $\phantom{+}0.9761$$(67)$\\
B1k3  &  $\phantom{+}0.00107$$(20)$  &  $-0.545$$(62)$  &  $-0.241$$(59)$  &  $\phantom{+}0.9701$$(47)$\\
B1k4  &  $-0.00057$$(19)$  &  $-0.486$$(68)$  &  $-0.316$$(51)$  &  $\phantom{+}0.9798$$(44)$\\
\hline
C1k2  &  $\phantom{+}0.00600$$(11)$  &  $-0.222$$(35)$  &  $-0.523$$(40)$  &  $\phantom{+}1.0489$$(29)$\\
C1k3  &  $-0.00109$$(11)$  &  $-0.304$$(54)$  &  $-0.698$$(50)$  &  $\phantom{+}1.0606$$(36)$\\
\hline
D1k2  &  $\phantom{+}0.00079$$(10)$  &  $-0.205$$(103)$  &  $-0.684$$(59)$  &  $\phantom{+}1.0849$$(52)$\\
D1k4  &  $-0.00007$$( 3)$  &  $-0.152$$(20)$  &  $-0.743$$(20)$  &  $\phantom{+}1.0885$$(11)$\\
\bottomrule
\end{tabular}

	\caption{
		Sea quark PCAC masses and estimators $\RAP$, $\Rm$ and $R_Z$ 
                for LCP-0 (i.e., at the unitary point) in the sector of 
                vanishing topological charge.
	}
	\label{tab:resAllEns_Q0}
\end{table}
\begin{table}%
	\renewcommand{\arraystretch}{1.25}
	\centering\small
	\begin{tabular}{cllll}
\toprule
ID & $m_{11}^{(\mathrm{all})}$ & $R_\text{AP}^{(\mathrm{all})}$ & $R_\text{m}^{(\mathrm{all})}$ & $R_Z^{(\mathrm{all})}$\\
\midrule
A1k1  &  $-0.00166$$(61)$  &  $-0.876$$(104)$  &  $\phantom{+}1.252$$(100)$  &  $\phantom{+}0.7458$$(55)$\\
A1k3  &  $\phantom{+}0.00262$$(130)$  &  $-0.587$$(133)$  &  $\phantom{+}1.139$$(102)$  &  $\phantom{+}0.7402$$(78)$\\
A1k4  &  $\phantom{+}0.00030$$(29)$  &  $-0.583$$(61)$  &  $\phantom{+}1.251$$(49)$  &  $\phantom{+}0.7485$$(36)$\\
\hline
E1k1  &  $\phantom{+}0.00308$$(22)$  &  $-0.652$$(89)$  &  $\phantom{+}0.272$$(51)$  &  $\phantom{+}0.8684$$(58)$\\
E1k2  &  $\phantom{+}0.00034$$(18)$  &  $-0.757$$(46)$  &  $\phantom{+}0.354$$(38)$  &  $\phantom{+}0.8715$$(32)$\\
\hline
B1k1  &  $\phantom{+}0.00562$$(14)$  &  $-0.375$$(37)$  &  $-0.286$$(31)$  &  $\phantom{+}0.9674$$(29)$\\
B1k2  &  $\phantom{+}0.00481$$(19)$  &  $-0.333$$(57)$  &  $-0.230$$(45)$  &  $\phantom{+}0.9677$$(44)$\\
B1k3  &  $\phantom{+}0.00164$$(16)$  &  $-0.575$$(43)$  &  $-0.136$$(42)$  &  $\phantom{+}0.9589$$(35)$\\
B1k4  &  $\phantom{+}0.00002$$(14)$  &  $-0.475$$(44)$  &  $-0.215$$(37)$  &  $\phantom{+}0.9712$$(29)$\\
\hline
C1k2  &  $\phantom{+}0.00619$$( 7)$  &  $-0.214$$(26)$  &  $-0.498$$(30)$  &  $\phantom{+}1.0461$$(25)$\\
C1k3  &  $-0.00086$$( 8)$  &  $-0.288$$(42)$  &  $-0.594$$(40)$  &  $\phantom{+}1.0547$$(26)$\\
\hline
D1k2  &  $\phantom{+}0.00084$$( 8)$  &  $-0.147$$(70)$  &  $-0.637$$(79)$  &  $\phantom{+}1.0837$$(47)$\\
D1k4  &  $-0.00002$$( 3)$  &  $-0.144$$(17)$  &  $-0.702$$(18)$  &  $\phantom{+}1.0867$$(11)$\\
\bottomrule
\end{tabular}

	\caption{
		Sea quark PCAC masses and estimators $\RAP$, $\Rm$ and $R_Z$ 
                for LCP-0 (i.e., at the unitary point) for all topological 
                sectors.
	}
	\label{tab:resAllEns_Qall}
\end{table}
\begin{table}%
	\renewcommand{\arraystretch}{1.25}
	\centering\small
	\begin{tabular}{cllll}
\toprule
ID & $m_\mathrm{PCAC}^{(0)}$ & $R_\text{AP}^{(0)}$ & $R_\text{m}^{(0)}$ & $R_Z^{(0)}$\\
\midrule
A1k1  &  $-0.00278$$(80)$  &  $-0.443$$(25)$  &  $\phantom{+}0.065$$(25)$  &  $\phantom{+}0.7805$$(52)$\\
A1k3  &  $\phantom{+}0.00079$$(118)$  &  $-0.349$$(34)$  &  $\phantom{+}0.011$$(28)$  &  $\phantom{+}0.7872$$(60)$\\
A1k4  &  $-0.00110$$(36)$  &  $-0.383$$(13)$  &  $-0.016$$(10)$  &  $\phantom{+}0.7889$$(27)$\\
\hline
E1k1  &  $\phantom{+}0.00262$$(26)$  &  $-0.329$$(16)$  &  $-0.264$$(14)$  &  $\phantom{+}0.8971$$(36)$\\
E1k2  &  $-0.00022$$(22)$  &  $-0.365$$(14)$  &  $-0.264$$(13)$  &  $\phantom{+}0.8994$$(28)$\\
\hline
B1k1  &  $\phantom{+}0.00549$$(21)$  &  $-0.197$$(15)$  &  $-0.439$$(12)$  &  $\phantom{+}0.9786$$(30)$\\
B1k2  &  $\phantom{+}0.00444$$(25)$  &  $-0.172$$(21)$  &  $-0.457$$(21)$  &  $\phantom{+}0.9845$$(43)$\\
B1k3  &  $\phantom{+}0.00107$$(20)$  &  $-0.249$$(18)$  &  $-0.454$$(17)$  &  $\phantom{+}0.9819$$(33)$\\
B1k4  &  $-0.00057$$(19)$  &  $-0.218$$(16)$  &  $-0.476$$(15)$  &  $\phantom{+}0.9882$$(30)$\\
\hline
C1k2  &  $\phantom{+}0.00600$$(11)$  &  $-0.099$$(10)$  &  $-0.586$$(11)$  &  $\phantom{+}1.0541$$(19)$\\
C1k3  &  $-0.00109$$(11)$  &  $-0.130$$(17)$  &  $-0.653$$(16)$  &  $\phantom{+}1.0623$$(27)$\\
\hline
D1k2  &  $\phantom{+}0.00079$$(10)$  &  $-0.088$$(30)$  &  $-0.670$$(20)$  &  $\phantom{+}1.0862$$(35)$\\
D1k4  &  $-0.00007$$( 3)$  &  $-0.069$$( 7)$  &  $-0.685$$( 7)$  &  $\phantom{+}1.0886$$( 8)$\\
\bottomrule
\end{tabular}

	\caption{
		Sea quark PCAC masses and estimators $\RAP$, $\Rm$ and $R_Z$ 
                for LCP-1 (i.e., at the partially-quenched point) in the 
                sector of vanishing topological charge.
	}
	\label{tab:resAllEns_Q0_Lx1}
\end{table}
\begin{table}%
	\renewcommand{\arraystretch}{1.25}
	\centering\small
	\begin{tabular}{cllll}
\toprule
ID & $m_\mathrm{PCAC}^{(\mathrm{all})}$ & $R_\text{AP}^{(\mathrm{all})}$ & $R_\text{m}^{(\mathrm{all})}$ & $R_Z^{(\mathrm{all})}$\\
\midrule
A1k1  &  $-0.00166$$(61)$  &  $-0.423$$(19)$  &  $\phantom{+}0.045$$(18)$  &  $\phantom{+}0.7805$$(34)$\\
A1k3  &  $\phantom{+}0.00262$$(130)$  &  $-0.324$$(25)$  &  $\phantom{+}0.014$$(18)$  &  $\phantom{+}0.7822$$(44)$\\
A1k4  &  $\phantom{+}0.00030$$(29)$  &  $-0.366$$( 9)$  &  $-0.013$$( 7)$  &  $\phantom{+}0.7864$$(18)$\\
\hline
E1k1  &  $\phantom{+}0.00308$$(22)$  &  $-0.337$$(12)$  &  $-0.251$$(10)$  &  $\phantom{+}0.8929$$(27)$\\
E1k2  &  $\phantom{+}0.00034$$(18)$  &  $-0.360$$(10)$  &  $-0.239$$( 9)$  &  $\phantom{+}0.8948$$(21)$\\
\hline
B1k1  &  $\phantom{+}0.00562$$(14)$  &  $-0.206$$( 9)$  &  $-0.445$$( 9)$  &  $\phantom{+}0.9779$$(18)$\\
B1k2  &  $\phantom{+}0.00481$$(19)$  &  $-0.197$$(14)$  &  $-0.440$$(13)$  &  $\phantom{+}0.9782$$(28)$\\
B1k3  &  $\phantom{+}0.00164$$(16)$  &  $-0.267$$(13)$  &  $-0.415$$(12)$  &  $\phantom{+}0.9730$$(25)$\\
B1k4  &  $\phantom{+}0.00002$$(14)$  &  $-0.230$$(10)$  &  $-0.439$$(11)$  &  $\phantom{+}0.9810$$(21)$\\
\hline
C1k2  &  $\phantom{+}0.00619$$( 7)$  &  $-0.104$$( 8)$  &  $-0.570$$( 8)$  &  $\phantom{+}1.0512$$(15)$\\
C1k3  &  $-0.00086$$( 8)$  &  $-0.137$$(13)$  &  $-0.613$$(14)$  &  $\phantom{+}1.0571$$(19)$\\
\hline
D1k2  &  $\phantom{+}0.00084$$( 8)$  &  $-0.081$$(22)$  &  $-0.646$$(22)$  &  $\phantom{+}1.0847$$(32)$\\
D1k4  &  $-0.00002$$( 3)$  &  $-0.070$$( 6)$  &  $-0.669$$( 6)$  &  $\phantom{+}1.0871$$( 8)$\\
\bottomrule
\end{tabular}

	\caption{
		Sea quark PCAC masses and estimators $\RAP$, $\Rm$ and $R_Z$ 
                for LCP-1 (i.e., at the partially-quenched point) for all 
                topological sectors.
	}
	\label{tab:resAllEns_Qall_Lx1}
\end{table}

\section{Correlations between the observables}
\label{app:correlations}

\begin{figure}[t]
	\centering
	\includegraphics[width=.75\linewidth]{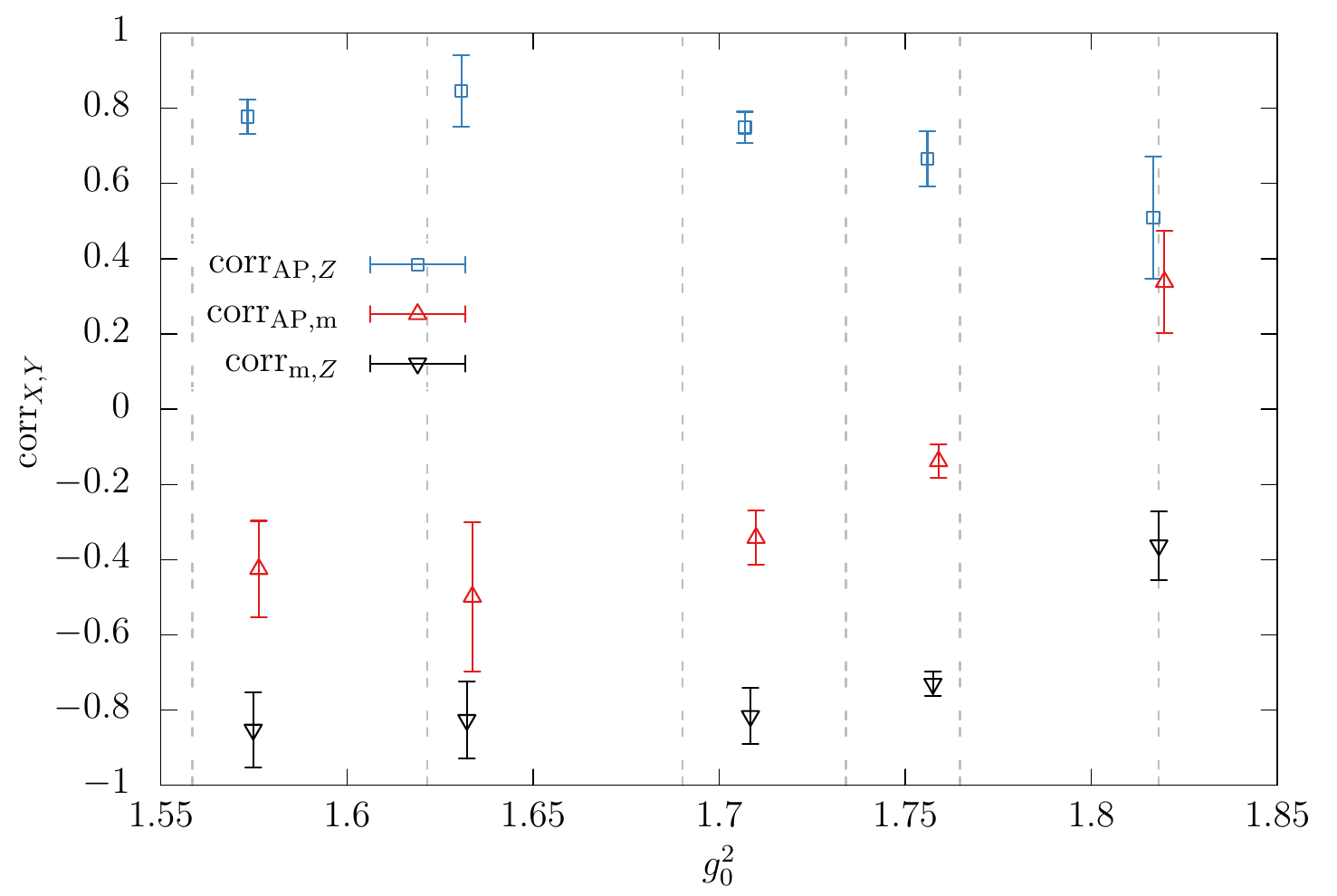}
	\caption{
		Correlations between the estimators $\RAP$, $\Rm$ and $R_Z$
                for the couplings used in our simulations. 
                The points are slightly shifted for visibility. 
                Vertical dashed lines indicate typical $g_0^2$-values of CLS 
                large-scale simulations.
	}
	\label{img:correlations}
\end{figure}

Since our final observables $\ba - \bp$, $\bm$ and $Z$ are determined on the 
same ensembles, we expect them to be correlated.
In table \ref{tab:restab_correlations} and figure \ref{img:correlations},
we give estimators for these correlations, i.e., 
\begin{align}
\mathrm{corr}_{X,Y}=
\frac{\mathrm{cov}_{X,Y}}{\sqrt{\mathrm{cov}_{X,X}\mathrm{cov}_{Y,Y}}}
\;,\qquad
X\neq Y\in\{{\rm AP}, {\rm m}, Z\} \;,
\end{align}
for LCP-0 at the five values of the coupling used in our calculations, and 
leave it to the reader to derive values at different couplings from them.

\begin{table}[t]%
	\centering\small
	\renewcommand{\arraystretch}{1.25}
	\begin{tabular}{cllllll}
\toprule
$\beta$ & $\mathrm{corr}_{\mathrm{AP},\mathrm{m}}$ & $\mathrm{corr}_{\mathrm{m},Z}$ & $\mathrm{corr}_{\mathrm{AP},Z}$\\
\midrule
$3.300$ & $\phantom{+}0.34(14)$ & $-0.36(10)$ & $\phantom{+}0.51(17)$ \\
$3.414$ & $-0.14( 5)$           & $-0.73( 4)$ & $\phantom{+}0.67( 8)$ \\
$3.512$ & $-0.34( 8)$           & $-0.82( 8)$ & $\phantom{+}0.75( 5)$ \\
$3.676$ & $-0.50(20)$           & $-0.83(11)$ & $\phantom{+}0.85(10)$ \\
$3.810$ & $-0.43(13)$           & $-0.85(11)$ & $\phantom{+}0.78( 5)$ \\
\bottomrule
\end{tabular}

	\caption{Correlations between the estimators $\RAP$, $\Rm$ and $R_Z$ 
                 for the $\beta$-values used in our simulations.
	}
	\label{tab:restab_correlations}
\end{table}

\clearpage
\small
\addcontentsline{toc}{section}{References}
\bibliography{}

\end{document}